\pdfoutput=1
\documentclass[pre,twocolumn,superscriptaddress]{revtex4}

\usepackage{amsmath}
\usepackage{graphicx}

\newcommand{\little}{2.in}
\newcommand{\medium}{\linewidth}
\newcommand{\figref}[1]{Fig.~\ref{#1}}
\newcommand{\sech}{{\text{sech}}}
\renewcommand{\eqref}[1]{Eq.~(\ref{#1})}

\newcommand{\F}{\mathcal{F}}

\newcommand{\EC}{\mathrm{Energy~curvature}}
\newcommand{\Damping}{\mathrm{Damping}}

\newcommand{\sol}{\mathrm{saddle}}
\newcommand{\saddle}{\mathrm{saddle}}
\newcommand{\straight}{\mathrm{straight}}
\newcommand{\hop}{\mathrm{hopping}}

\newcommand{\D}{D}
\newcommand{\eff}{\mathrm{eff}}

\begin{document}

\title{Nucleation at the DNA supercoiling transition}
\author{Bryan C.~Daniels}
\affiliation{Department of Physics, 
Laboratory of Atomic and Solid State Physics, Cornell University, 
Ithaca, NY 14853}
\affiliation{Santa Fe Institute, Santa Fe, NM 87501}
\author{James P.~Sethna}
\affiliation{Department of Physics, 
Laboratory of Atomic and Solid State Physics, Cornell University, 
Ithaca, NY 14853}
\begin{abstract}
Twisting DNA under a constant applied force reveals a thermally activated 
transition into a state with a supercoiled structure known as a plectoneme.
Using transition state theory, we predict the rate of this plectoneme 
nucleation to be of order $10^4$~Hz.  We reconcile this with experiments that
have measured hopping rates of order 10~Hz by noting that the viscous drag on the
bead used to manipulate the DNA limits the measured rate.  We 
find that the intrinsic bending caused by disorder in the
base-pair sequence is important for understanding the free
energy barrier that governs the transition.  Both analytic and
numerical methods are used in the calculations.  We provide extensive
details on the numerical methods for simulating the elastic rod
model with and without disorder.
 
\end{abstract}
\maketitle

\section{Introduction}

\begin{figure}
\centering
\includegraphics[width=\medium]{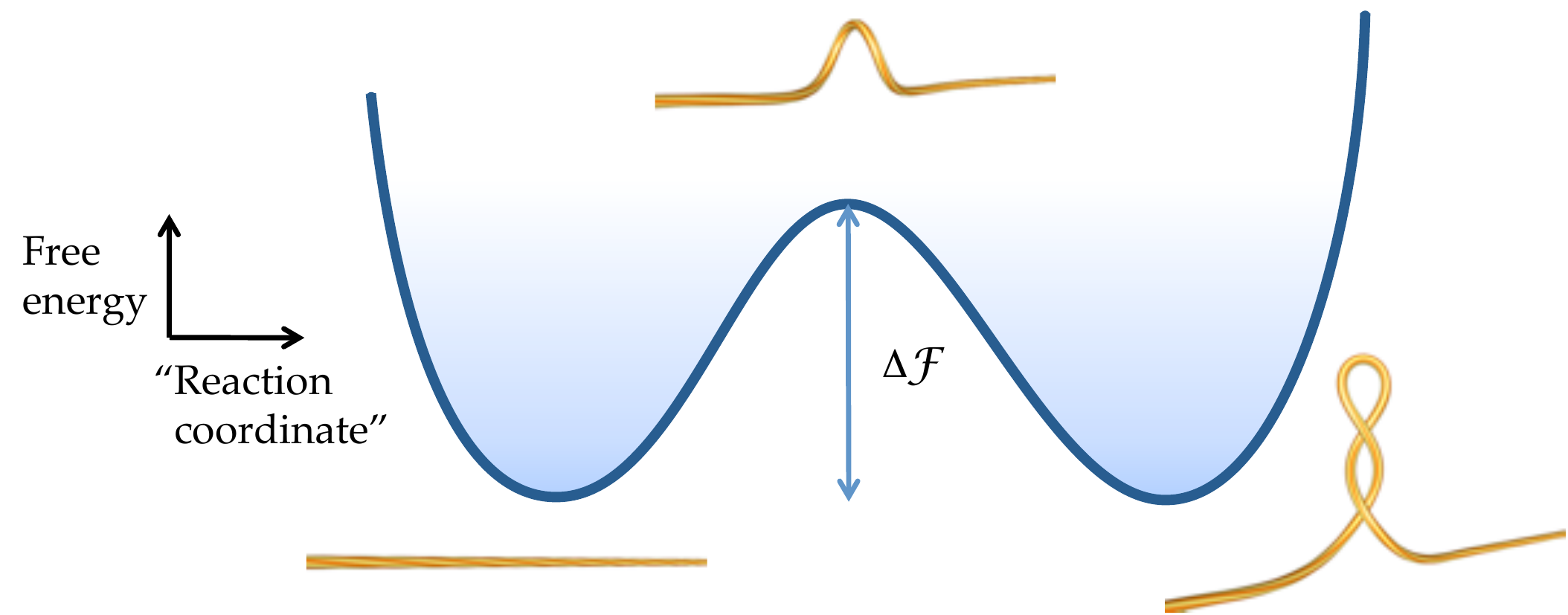}
\caption{ (Color online) The free energy double-well: \label{DoubleWellCartoon}%
A schematic one-dimensional representation of the two 
metastable states and the saddle point that separates them.  
}
\end{figure}

When overtwisted, DNA forms supercoiled structures known as
plectonemes (as seen in the lower right of 
\figref{DoubleWellCartoon}), familiar from phone cords and 
water hoses.
Single molecule experiments commonly hold a molecule of DNA
under constant tension and twist one end; the appearance of
a growing plectoneme can be thought of as the nucleation
of a new phase that can store some of the added twists
as writhe.  
Recent experiments that hold DNA near this 
supercoiling transition have shown that it
is not initiated by a linear instability (as it is
in macroscopic objects), 
but is rather an equilibrium transition between two 
metastable states, with and without a plectoneme.
These states are separated by
a free energy barrier that is low enough to allow thermal
fluctuations to populate the two states, but high enough
that the characteristic rate of hopping is only about 10~Hz.
Two experimental groups, using different methods to manipulate
the DNA (one, an optical trap \cite{ForDeuShe08}; the other, 
magnetic tweezers \cite{BruLuzKla10}), have observed this 
nucleation at the transition and reported similar qualitative 
and quantitative results.  

Understanding the rate of plectoneme nucleation at the 
supercoiling transition is a useful goal for both biology 
and physics.  First, the biological function of DNA is
tied to its microscopic physical characteristics, and
plectoneme nucleation is sensitive to many of these. 
The microscopic dynamics of DNA in water is one such 
factor: while often theorized as a cylindrical rod 
in a viscous liquid, these dynamics have not been well studied 
experimentally.  The nucleation rate is also 
sensitive to the intrinsic bend disorder present 
in a given DNA sequence, potentially providing information 
to clarify the degree of bend disorder, which is debated
in current literature \cite{BedFurKat95,VolVol02,TriTanHar87,Nel98}.  
Popular elastic rod models for
DNA can also be tested.
Second, DNA supercoiling provides to
physics a unique testing ground for theories of
thermal nucleation.  The
theory of thermal nucleation in spatially extended 
systems (critical droplet theory) was essentially proposed 
in its current form by Langer in the
1960's~\cite{Lan68,Lan69} 
(see H{\"a}nggi~\cite[section IV.F]{HanTalBor90}, and also 
Coleman~\cite{Col77} for the corresponding 
`instanton' quantum tunneling analogue.)
Experimental validation of 
these theories has been difficult in bulk 
systems, however, for reasons that DNA nucleation neatly
bypasses.  (a)~The nucleation rate in most systems is 
partly determined by the atomic-scale surface tension; in 
DNA the continuum theory describes the entire nucleation 
process.  (b)~Nucleation in bulk phases is rare (one 
event per macroscopic region per quench), and hence 
typically has a high energy barrier. Small estimation 
errors for this barrier height typically hinder 
quantitative verification of the (theoretically interesting) 
prefactor. In single-molecule DNA experiments, the 
plectoneme nucleation barrier is only a few $k T$, 
and indeed rather short segments of DNA exhibit multiple 
hops over the barrier --- allowing direct measurements 
of the transition rate in equilibrium.
(c)~Nucleation 
in bulk phases is normally dominated by disorder (raindrops 
nucleate on dust and salt particles); in DNA the likely dominant 
source of disorder (sequence-dependence) is under the 
experimentalist's control.

As an illustrative example, consider the classic early study of supercurrent
decay in thin wires \cite{LanAmb67,McCHal70}. Here (a)~the superconductor (like DNA)  
is well described by a continuum theory (Ginzburg-Landau theory) because
the coherence length is large compared to the atomic scale, but
(c)~the rate for a real, inhomogeneous wire will strongly depend on,
for example, local width fluctuations. Finally, (b)~the rate of 
nucleation is so strongly dependent on experimental conditions that an early
calculation~\cite{LanAmb67} had an error in the prefactor of a factor 
of $10^{10}$~\cite{McCHal70},
but nonetheless still provided an acceptable agreement with experiment.

Reaction-rate theory predicts a rate of plectoneme
nucleation related
to the energy barrier between the two states.  
We perform a full calculation of this energy barrier and
of the rate prefactor, 
including hydrodynamics, entropic
factors, and sequence-dependent intrinsic bend disorder, to determine which
effects contribute to this slow rate.  We calculate a rate
of order $10^4$~Hz, about 1000 times faster than measured
experimentally.  The discrepancy can be attributed to a slow timescale
governing the dynamics of the measurement apparatus.
The experiments measure the extension by monitoring the position 
of a large bead connected to one end of the DNA, and its dynamics 
are much slower than that of the DNA strand --- thus the bead hopping rate
that the experiments measure is much slower than the plectoneme nucleation 
hopping rate that we calculate.  Future experiments may be able to 
slow the plectoneme nucleation rate enough that the bead dynamics is
unimportant, such that the bead motion would reveal the dynamic 
characteristics of the DNA itself.


We begin in section~\ref{nucleationRateSection} with the 
nucleation rate calculation in the 
absence of disorder: \ref{saddlePointEnergySection} gives the 
saddle-point energy, \ref{transitionStateBasicIdeaSection} 
gives the 
technique for calculating the prefactor, \ref{DynamicsSection} 
overviews the 
dynamics of DNA in water, and \ref{FullDerivationSection} gives 
the transition-state 
theory calculation in full. Section~\ref{undisorderedResultsSection} 
presents the results of 
the undisordered calculation, with qualitative explanations of 
the magnitudes of the various terms, and shows that the results 
are incompatible with the experiments. This motivates our discussion 
of base-pair disorder in section~\ref{hoppingDisorderSection}, 
where we estimate the 
disorder-renormalization of the elastic constants in 
section~\ref{renormalizationSection}, 
and formulate and calculate the rates in 
sections~\ref{rateWithDisorderSection} and \ref{disorderResultsSection}. 
We conclude in section~\ref{conclusionSection}. 

We draw the reader's 
attention in particular 
to the appendices, where substantive, general-purpose results are 
presented for numerical discretization and calculations with the 
elastic rod model. 
Appendix~\ref{energyCalculationSection} reformulates the 
Euler-angle description in terms of more geometrically natural 
rotation matrices. Appendix~\ref{transitionStateCalculationSection} 
explains how to transform a DNA with 
$N$ segments from the $3N$-dimensional Euler-angle or rotation-matrix 
space to the $4N$-dimensional $(x,y,z,\phi)$ space of the natural 
dynamics, and the Jacobians needed to transform path integrals 
over the latter into path integrals over the former. 
Appendix~\ref{numericalDetailsSection} 
discusses the discretization and the rotation-invariant forms for 
bend and twist in terms of rotation matrices. Finally, 
Appendix~\ref{disorderAppendix} 
provides our numerical implementation of randomly-bent DNA, 
mimicking the effects of a random base-pair sequence.

\section{Nucleation rate calculation}
\label{nucleationRateSection}

\subsection{Saddle point energetics}
\label{saddlePointEnergySection}

To understand the dynamics of nucleation, we must first find 
the saddle point DNA configuration that
serves as the barrier between the stretched state and the
plectonemic state.  We model the DNA as an inextensible 
elastic rod, with total elastic energy
\begin{equation}
\label{EqElastic}
E_{\mathrm{elastic}} = 
	\int_{0}^{L} ds [ \frac{B}{2} \beta(s)^2 + \frac{C}{2} \Gamma(s)^2 ],
\end{equation}
where $s$ is arclength along the rod, 
$\beta$ and $\Gamma$ are the local bend and twist deformation angles,
respectively, and $L$ is the contour length of the rod 
(for more details, see Appendix~\ref{elasticRodModel}; the
dynamical equations of motion will be discussed in 
section~\ref{DynamicsSection}).
Fain and 
coworkers used variational techniques to characterize the
extrema of this elastic energy functional \footnote{
    They take the infinite length limit, using fixed 
    linking number and force boundary conditions.
};
this revealed a ``soliton-like excitation'' as the 
lowest-energy solution with nonzero writhe 
\cite{FaiRudOst96}.  They found that the soliton's
energy differed from that of the straight state by a finite
amount in the infinite-length limit.  This soliton state, depicted
at the top of the barrier in \figref{DoubleWellCartoon},
is the one we identify as the saddle configuration.

The shape of the saddle configuration is controlled by the
bend and twist elastic constants $B$ and $C$, as well as the torque
$\tau$ and force $F$ applied as boundary conditions.
Defining the lengths
\begin{align}
a &\equiv \sqrt{\frac{B}{F}} \\
b &\equiv \frac{2B}{\tau} \\
\ell &\equiv ( a^{-2} - b^{-2} )^{-1/2},
\label{ellEqn}
\end{align}
we can write the Euler angles characterizing the saddle 
configuration as \cite{FaiRudOst96}\footnote{
We find that the analytical saddle configurations
from Eqs.~(\ref{SaddleEulerAngles}) do not produce states with precisely
zero forces in our numerical calculations --- we attribute this to the 
effects of discretization.  We therefore
first minimize the forces using the same procedure used to find 
saddle states with intrinsic bending disorder, 
described in section~\ref{findingSaddlePointsSection}.
}
\begin{align}
\nonumber
\cos \theta(s) &= 1 - 2\left(\frac{a}{\ell}\right)^2 \left[ 1 
                - \tanh^2\left(\frac{s}{\ell}\right) \right] \\
\label{SaddleEulerAngles}
\phi(s) &= \quad \quad \:   + \: \frac{s}{b} 
        + \tan^{-1}\left[ \frac{b}{\ell} \tanh\left(\frac{s}{\ell}\right) \right] \\
\nonumber
\psi(s) &= \frac{s}{C/\tau} - \frac{s}{b}
        + \tan^{-1}\left[ \frac{b}{\ell} \tanh\left(\frac{s}{\ell}\right) \right],
\end{align}
where $s$ is arclength along the DNA backbone.

For the experiments in which we are interested, 
$L/\ell~\gg~1$ (the soliton ``bump'' is much smaller than the
length of the DNA), such
that we can safely remove the soliton from the infinite length
solution and still have the correct boundary conditions 
[$\theta(\pm L/2) = 0$, 
such that the tangent vector $\hat t$ points
along the $z$ axis at the ends of the DNA].  In this case, the linking number in the saddle
state is given by \cite{FaiRudOst96}
\begin{align} \nonumber
K_s &= \frac{1}{2\pi} 
    \left [ \phi(L/2) - \phi(-L/2) + \psi(L/2) - \psi(-L/2) \right] \\
\label{saddleLinkEqn}
    &= \frac{\tau L}{2\pi C} + W_s(\tau),
\end{align}
where we have used \eqref{SaddleEulerAngles} and separated the linking 
number into twist (the first term) and writhe:
\begin{equation}
W_s(\tau) = \frac{2}{\pi} 
    \tan^{-1}\left[ \frac{b}{\ell} \tanh\left(\frac{L}{2\ell}\right) \right].
\end{equation}
To find the saddle configuration's torque at the supercoiling 
transition \footnote{
    Note that this is not the same as the torque before or after
    the transition.
},
we numerically solve \eqref{saddleLinkEqn} for $\tau$ using the 
experimentally observed critical linking number $K_s^*$ \footnote{
    We could
    alternatively use $K_s^* = 8.1$ from the theory in 
    Ref.~\cite{DanForShe09}, but finding the transition
    involves the complications of the full plectonemic state, including
    entropic repulsion, that are less well-understood than the elastic
    properties.  Using this alternative $K_s^*$ increases $\Delta E$ by
    about $1~kT$, which does not significantly alter our conclusions.
}.

The energy barrier $\Delta E$ is the difference in 
elastic energy between the saddle and straight states at the same 
linking number $K_s$ \footnote{
    What about the barrier from the plectoneme to the saddle point? 
    Since the plectoneme is stabilized by self-repulsion 
    (electrostatic and entropic), analytic calculations are more 
    difficult. But at forces and torques where the plectoneme is in 
    coexistence with the straight state, the total plectoneme free 
    energy is equal to that of the straight state, and hence the 
    free energy barriers are the same.
}. We find \footnote{
    Taking $L \rightarrow \infty$, this agrees with 
    Eq.~(19) of Ref.~\cite{FaiRudOst96} if their $F/L$ is replaced 
    by $\sqrt{2FB}$.
}
\begin{equation}
\label{EnergyBarrierEqn}
\Delta E = \frac{8B}{\ell} \tanh \left( \frac{L}{2\ell} \right)
    - 2\pi W_s \left( \tau + \frac{\pi C}{L} W_s \right).
\end{equation}
Inserting the experimental values listed in 
Table~\ref{parameterTable} into \eqref{EnergyBarrierEqn}, 
we calculate an energy barrier 
\begin{equation}
\Delta E = 5.5~kT.
\end{equation}
This barrier would seem surprisingly small considering 
that typical atomic rates are on the order of $10^{13}$~Hz:
using this for the attempt frequency in an activated rate 
would give $10^{13}e^{-5.5} = 10^{11}$~Hz
for the hopping rate.  The next sections present a more careful
calculation, which shows that the timescale for motion over the
barrier is in fact many orders of magnitude smaller than $10^{13}$~Hz 
due to the larger length scales involved (and even smaller when we 
calculate the bead hopping rate), but also that
the entropy from multiple available nucleation sites 
significantly lowers the barrier.

\begin{table*}
\begin{minipage}{\textwidth}
\centering
\caption[]{Parameter values for nucleation rate calculation.}
\begin{tabular}{|l|l|l|}
\hline
\label{parameterTable}
Symbol  & Description                           & Value         \\
\hline
$B$     & bend elastic constant                 & (43 nm) $kT$ \cite{ForDeuShe08} \\
$C$     & twist elastic constant                & (89 nm) $kT$ \cite{ForDeuShe08} \\
$F$     & applied force                         & 1.96 pN \cite{ForDeuShe08}      \\
$kT$    & thermal energy at 23.5$^\circ$C       & 4.09 pN nm    \\
$L$     & basepair length of DNA strand         & 740 nm \cite{ForDeuShe08}       \\
$K_s^*$ & critical linking number               & 8.7 \cite{ForDeuShe08}          \\
$\tau$  & saddle point torque [\eqref{saddleLinkEqn}] \footnotemark[1]        
                                                & 25 pN nm      \\
$\ell$  & soliton length scale [\eqref{ellEqn}] \footnotemark[1] 
                                                & 13 nm         \\
$R$     & bead radius                           & 250 nm        \\
$\eta$  & viscosity of water at 23.5$^\circ$C   & $9.22 \times 10^{-10}$ pN s / nm$^2$ \\
$N$     & number of segments                    & 740           \\
$d$     & length of segment                     & 1 nm          \\
$r_{D}$ & DNA hydrodynamic radius               & 1.2 nm        \\
$\zeta$ & translation viscosity coeff. [\eqref{zetaEqn}]    
                                                & $1.54 \times 10^{-9}$ pN s / nm$^2$  \\
$\lambda$& rotation viscosity coeff. [\eqref{lambdaEqn}]      
                                                & $1.67 \times 10^{-8}$ pN s           \\
\hline
\end{tabular}
\centering

\footnotemark[1]
    These values have been calculated using $B_m = B$, that is, for
    disorder $D = 0$.  See \eqref{BmEqn}.
\end{minipage}
\end{table*}


\subsection{Transition state theory: the basic idea}
\label{transitionStateBasicIdeaSection}


When, as in our case, the energy barrier is much larger than the 
thermal energy $kT$, the rate of nucleation is suppressed by 
the Arrhenius factor $\exp{(-\Delta E/kT)}$.  Going beyond this
temperature dependence to an estimate of the
full rate, however, requires a more detailed calculation.
We will follow the prescription from Kramers' spatial-diffusion-limited
reaction-rate theory \cite{HanTalBor90} to
calculate the rate of hopping.  The requirements are that 
(1) the timescales involved in motion within the two metastable
wells are much faster than the timescale of hopping, 
and (2) (for Kramers' ``spatial-diffusion-limited'' 
theory) the system is overdamped, in the
sense that the ratio of damping strength to the rate of
undamped motion over the barrier top is large.  We check
that these requirements are met for the intrinsic DNA
nucleation rate after 
the calculation in section~\ref{FullDerivationSection}.

Under these two conditions, Kramers' reaction-rate theory tells us that the 
rate of hopping over the barrier is controlled only by the rate of
motion through the ``narrow pass'' at the top of the barrier, since it
is much slower than any other timescale in the system.   This 
means that the hopping rate should be the characteristic rate of motion
across the barrier top times the 
probability of finding the system near the barrier top, which, in terms
of the curvature in the unstable direction away from the saddle point,
we can write schematically as
\begin{align} \nonumber
k_\hop 
    &= \left(\mathrm{Characteristic~rate~of~motion~at~barrier~top}\right) \\
    &\quad \times(\mathrm{Prob.~of~being~at~top})
    \label{simplerRateEquation}  
    \\ \label{simpleRateEquation} 
    &= \left(\frac{\EC}{\Damping}\right)
    (\mathrm{Prob.~of~being~at~top}).
\end{align}
 
It is important to note that, in current experiments, the measuring
apparatus violates condition~(1) above.  The measurement of the
extension is only an indirect readout of the configurational 
state of the DNA --- it is a measure of the position 
of a large bead connected to one end of the DNA strand.  If the
bead has much slower dynamics than the DNA, then it will set the 
characteristic rate of motion in \eqref{simplerRateEquation}.
In section~\ref{HistogramSection}, we will find that this is the case for
the experimental numbers we use.  Therefore, the rate we will 
calculate is a plectoneme nucleation hopping rate that is not
the same as the (slower) bead hopping rate.  We will find 
also that future
experiments may be able to measure the underlying plectoneme
nucleation hopping rate by testing regimes where the bead hopping
rate is not limited by the bead dynamics.

\subsection{Dynamics of DNA in water: the diffusion tensor}
\label{DynamicsSection}

To find the rate of motion over the barrier top, we need to know
the microscopic dynamics.  We will be treating the DNA strand as a
series of cylindrical segments, parametrized by the Cartesian 
coordinates $(x_n,y_n,z_n)$ of one end of each segment plus the 
Euler angle $\psi_n$ that controls local twist 
(see section~\ref{choosingCoordsSection}
for a discussion about the choice of coordinates). 
Assuming overdamped motion such that we can
neglect inertial terms, we will write the equations of
motion in the form [with $\vec r_n = (x_n,y_n,z_n,\psi_n)$, 
$i,j$ labeling coordinates, and
$m,n$ labeling segments] \cite{HanTalBor90}
\begin{equation}
\label{equationsOfMotion}
\frac{d r_{m i}}{dt} = - M_{m i, n j} \frac{d E}{d r_{n j}};
\end{equation}
$M$ is the diffusion tensor, which transforms forces 
to velocities.

The simplest diffusion tensor produces motion proportional to
the local forces, making it diagonal in segment number $n$ and
coordinate $i$:
\begin{equation}
\label{MDiagonal}
   M^{\mathrm{diagonal}}_{m i, n j} =  \left\{
     \begin{array}{ll}
       \frac{1}{d \zeta} \, \delta_{m n} \, \delta_{i j}
                                & \mathrm{for}~i,j \in \{x,y,z\} \\
       \frac{1}{d \lambda} \, \delta_{m n} & \mathrm{for}~i=j=\psi
     \end{array}
   \right.
\end{equation} 
The viscous diffusion constants are set so they reproduce the
known diffusion constant for a straight cylinder of length $B/kT$ 
(the bending persistence length) and radius $r_{D} = 1.2$~nm:
\cite{BalKouMah04,Cox70}
\begin{align}
\label{zetaEqn}
\zeta &= \frac{2\pi \eta}{\ln{(B/(kT \, r_{D}))}} \\
\label{lambdaEqn}
\lambda &= 2\pi \eta \, r_{D}^2,
\end{align}
where the viscosity of water $\eta = 9.22 \times 10^{-10}$~pN\,s/nm$^2$ 
at the experimental temperature of $23.5^\circ$ C.

It is important to note that the DNA in this experiment is 
attached to a large bead
(with radius $R \approx 250$~nm) that must also be pulled through the 
water during the transition \footnote{
    It is also known that the presence of a surface (the glass plate to which
    the DNA is attached) leads to an enhanced
    hydrodynamic diffusion on the bead that depends on the distance
    $z$ from the surface; for motion perpendicular to the 
    surface \cite{CruKosSei07},
    $\zeta_{\mathrm{eff}} = \zeta [1 + R/z + R/(2R+6z)]$.  For our 
    typical $z$ and $R$, this increases $\zeta$ for the
    bead by a factor of about 1.6, which does not significantly change our
    conclusions, so we choose not to include this correction.
}.  We take this into account by setting
the translational diffusion constant for the final segment in the chain
according to Stokes' Law: 
\begin{equation}
M_{N i, N j} = \delta_{i j}/(6\pi \eta R) ~~~~ \mathrm{for}~i,j \in \{x,y,z\}.
\end{equation}

\begin{figure}
\centering
\includegraphics[width=\medium]{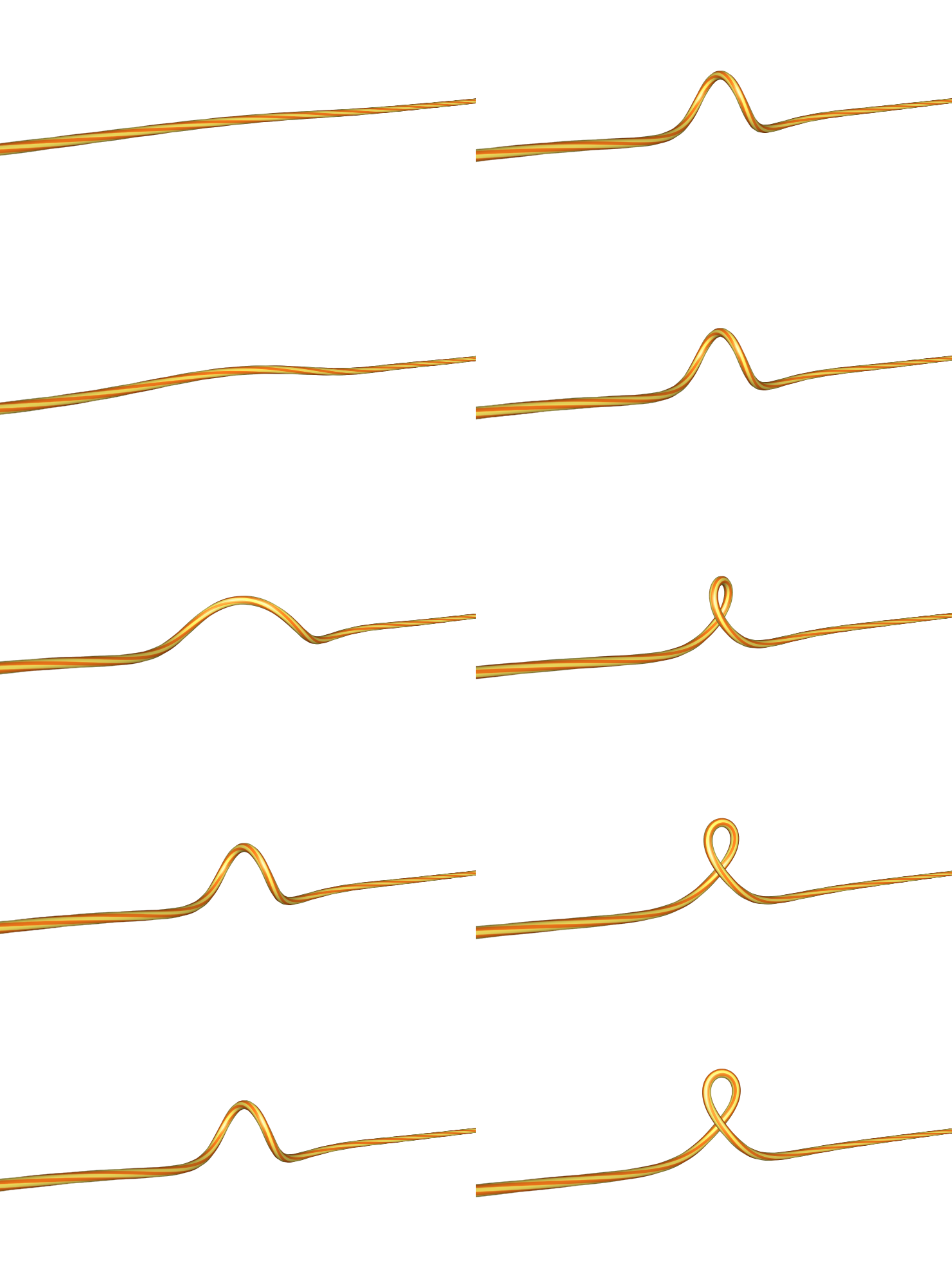}
\caption{ (Color online)
Snapshots along the transition path. \label{UnstableModePictureTiled}%
After perturbing the saddle 
state along the unstable direction $u$ (bottom left and top right snapshots), 
we integrate the 
equations of motion in \eqref{equationsOfMotion} to follow the unstable 
dynamics into the two metastable wells.  Here, we use Rotne-Prager
dynamics with $\D=0$, the timestep between frames is $1.5 \times 10^{-6}$ s,
and other parameters are as given in Table~\ref{parameterTable}.
}

\end{figure}

Hydrodynamic effects may also be important, which introduce interactions
between segments: as segments move, they
change the velocity of the water around them, 
and this change propagates to change the viscous force felt by other
nearby segments.  Following Ref.~\cite{KleMerLan98}, we incorporate
hydrodynamic effects by using a Rotne-Prager tensor for the translational
diffusion, modeling the
strand as a string of beads \footnote{
    The Rotne-Prager tensor (also known as the Rotne-Prager-Yamakawa
    tensor \cite{DoyUnd04}) is a regularized version of the 
    Kirkwood-Riseman diffusion tensor (also known as the Oseen-Burgers
    tensor \cite{DoyUnd04}) that is modified at short distances
    such that it becomes positive definite, producing stable 
    dynamics \cite{RotPra69}.
}:
\begin{widetext}
\begin{equation}
\label{MRotnePrager}
  M^{\mathrm{Rotne-Prager}}_{m i, n j} = \left\{
    \begin{array}{ll}
      D_0 \frac{3a}{4r_{mn}} \left[ \delta_{i j} 
        + \frac{r_{mn,i} r_{mn,j}}{r_{mn}^2} 
        + \frac{2a^2}{3r_{mn}^2} \left( \delta_{i j} 
                    - 3 \frac{r_{mn,i} r_{mn,j}}{r_{mn}^2} \right) \right] \\
        \quad \quad \quad \mathrm{for}~r_{mn} \geq 2a,~m \neq n \\
      D_0 \left[ \left( 1 - \frac{9}{32} \frac{r_{mn}}{a} \right) \delta_{i j}
        + \frac{3}{32} \frac{r_{mn,i} r_{mn,j}}{a\,r_{mn}} \right] \\
        \quad \quad \quad \mathrm{for}~r_{mn} \leq 2a,~m \neq n \\
      D_0 \, \delta_{ij}~\,\mathrm{for}~ m = n
    \end{array}
  \right.
\end{equation}
for $i,j \in \{x,y,z\}$, where $D_0 = (6\pi \eta a)^{-1}$ 
and $a$ is an effective 
bead radius chosen such that a
straight configuration of Kuhn length $L_K = 2B/kT$ 
(with a number of beads $L/L_K$) has the same
total diffusion constant as a cylinder of length $L$ and
radius $r_{D}$
(see \cite{KleMerLan98}).  With the parameters in 
Table~\ref{parameterTable}, we use $a = 0.98$~nm.

\subsection{Transition state theory: full calculation}
\label{FullDerivationSection}

In the full multidimensional space inhabited by our model, the saddle
configuration will have a single unstable direction that locally
defines the ``reaction coordinate'' depicted in \figref{DoubleWellCartoon}.
The direction of the unstable mode can be found numerically by locally
solving the equations of motion \eqref{equationsOfMotion}.  First, the local
quadratic approximation to the energy is provided by the Hessian
\end{widetext}
\begin{equation}
H_{mi,nj} = \frac{d^2 E}{d r_{mi} d r_{nj}},
\end{equation}
where the derivatives are taken with respect to the unitless variables 
$\vec r = {x/\ell_0,y/\ell_0,z/\ell_0,\psi}$ 
(where $\ell_0$ is an arbitrary length scale \footnote{
    Choosing these units for our variables makes the path integral partition 
    function unitless:
    $
    Z = \int \prod_n \frac{dx_n dy_n dz_n d\psi_n}{l_0^{3}}     
        e^{-E(\{x,y,z,\psi\})/kT}.
    $
}; see also section \ref{choosingCoordsSection}). 
Inserting the quadratic form defined by $H$ at the saddle point 
into \eqref{equationsOfMotion}, we 
then diagonalize the matrix
$M H_{\saddle}$ to find the dynamical normal modes of the system; the single
mode $u$ with a negative eigenvalue $-\lambda_b$ is the unstable mode at the
top of the barrier:  
\begin{equation}
M_{li,mj} H^{\saddle}_{mj,nk} \, u_{nk} = -\lambda_b \, u_{li},
\end{equation}
and $\lambda_b$ defines the characteristic rate of 
\eqref{simpleRateEquation}.  We have checked that we find the
correct saddle configuration and unstable mode $u$ by perturbing forward
and backward along $u$ and numerically integrating the dynamics 
of \eqref{equationsOfMotion} --- one case ends in the straight
state well and the other in the plectonemic state well.  This generates
the transition path connecting the two wells, as illustrated 
in \figref{UnstableModePictureTiled}.


To find the probability of being near the top of the barrier
in the multidimensional case, we need to know not only the
energy barrier $\Delta E$, but
also the entropic factors coming from the amount of narrowing in
directions transverse to the transition path \footnote{
    The rate will be slower if the system must traverse a more narrow 
    ``pass'' at the saddle point.
},
which are controlled by the remaining eigenvalues of $H_{\saddle}$ and
the Hessian $H_{\straight}$ of the straight state.
The full result from spatial-diffusion limited multidimensional
transition state theory is (to lowest order in $kT$) \cite{HanTalBor90}
\begin{equation}
\label{simpleRateDets}
k_\hop = \frac{\lambda_b}{2\pi} 
    \sqrt{ \frac{\det H_{\straight}/(2\pi k T)}{|\det H_{\sol}/(2\pi k T)|} }
    e ^ {-\Delta E/kT};
\end{equation}
or, in terms of the eigenvalues of each Hessian,
\begin{equation}
\label{simpleRateEigenvalues}
k_\hop = \frac{\lambda_b}{2\pi} 
    \sqrt{ \frac{\prod_{i=1}^{4(N-2)} \lambda_i^{\straight}/(2\pi k T)}
                {|\prod_{i=1}^{4(N-2)} \lambda_i^{\sol}/(2\pi k T)|} }
    e ^ {-\Delta E/kT}.
\end{equation}
These are correct if each eigenvalue is sufficiently large such that 
the local quadratic form is a good approximation where $E \lesssim kT$.  


In our case, we must deal separately with the two zero modes due to invariance 
with respect to location $s_s$ and rotation angle 
$\rho_s$ of the saddle configuration's bump.  Extracting these directions 
from the saddle integral, we have
\begin{widetext}
\begin{align} \nonumber
k_\hop &= \frac{\lambda_b}{2\pi} 
    \left(\int J_s J_\rho \frac{ds_s}{\ell_0} d\rho_s \right)
    \sqrt{ \frac{\prod_{i=1}^{4(N-2)} \lambda_i^{\straight}/(2\pi k T)}
                {|\prod_{i=3}^{4(N-2)} \lambda_i^{\sol}/(2\pi k T)|} }
    e ^ {-\Delta E/kT} \\
 &= \frac{\lambda_b}{2\pi} 
    \left(2\pi \frac{L}{\ell_0} J_s J_\rho \right)
    \frac{1}{2\pi kT}\sqrt{ \frac{\det H_{\straight}} {|\det' H_{\sol}|} }
    e ^ {-\Delta E/kT}, \label{noDisorderRateEqn}
\end{align}
\end{widetext}
where the Jacobians $J_s = | d\vec r_s / ds_s |$ and 
$J_\rho = | d\vec r_s / d\rho_s |$, and $\det '$ represents the 
determinant without the two zero modes (but including the unstable mode).
Numerically, $J_s$ and $J_\rho$ are calculated using the known forms for
derivatives of the saddle point's Euler angles $\alpha_s$ 
[Eqs.~(\ref{SaddleEulerAngles})] with respect to $s_s$ and $\rho_s$:
$J_s = | [J^T(\vec r_s)]^{-1} d\vec \alpha_s/ds_s |$ and 
$J_\rho = | [J^T(\vec r_s)]^{-1} d\vec \alpha_s/d\rho_s |$, where
$J$ is defined in \eqref{Jacobian}.

We can now check that we meet the requirements for using Kramers'
theory set out in section~\ref{transitionStateBasicIdeaSection}.  
First, we check condition (1) by looking at the smallest nonnegative eigenvalues
of $MH$.  The slowest mode is transverse motion with wavelength
$2L$, and since the bead has much larger viscous drag than the 
rest of the DNA chain, it sets the damping for this motion;
this produces a frequency $F/(6\pi \eta R L) 
\approx 600$~Hz.  The other modes all have frequencies of order
$10^4$~Hz or faster.  We will find that the calculated Kramers rate of hopping
lies between these two timescales --- this means that, while the bead motion is too
slow to follow the fast hopping, Kramers' theory should correctly give
the plectoneme nucleation hopping rate for a fixed bead position.
We can check (2) by comparing the characteristic rate for undamped
barrier motion to the characteristic damping rate.  In the
spirit of section~\ref{OrderOfMagnitudeSection}, we are dealing with a 
portion of DNA of
length $\ell_B$, such that it has a mass $\mu \ell_B$, where 
$\mu = 3.3 \times 10^{-21}$~g/nm is the linear mass density
of DNA \cite{Mul06}, and the energy curvature at the barrier top
is on the order of $\pi^4 B/ \ell_B^3$.  Then the damping coefficient is 
$\zeta/\mu = 5 \times 10^{11}~\mathrm{s}^{-1}$, and the 
rate for undamped barrier motion is 
$\sqrt{ (\pi^4 B/ \ell_B^3) / (\mu \ell_B) } = 4 \times 10^8~\mathrm{s}^{-1}$.
Since the ratio of these values is much greater than one, 
we are firmly in the overdamped regime, and Kramers' rate theory
applies \cite{HanTalBor90}.


\section{Initial results and order of magnitude checks}
\label{undisorderedResultsSection}

\subsection{Initial results}


We calculate the rate in \eqref{noDisorderRateEqn} using numerical
methods described in the Appendices.  
We will quote the results of the calculation by looking 
individually at the factors that contribute to the rate.  Writing
the rate in various simple forms,
\begin{align}
\label{effectiveFreeEnergyEqn}
k_\hop &= \frac{\lambda_b}{2\pi} e^{-\Delta E/kT + S/k} \\
\label{effectiveFreeEnergyEqn2}
  &= \frac{\lambda_b}{2\pi} e^{-\Delta \F/kT};
\end{align}
$S$ encapsulates the entropic factors coming
from fluctuations in the straight and saddle configurations, and
the effective free energy barrier $\Delta \F = \Delta E-TS$ 
provides the relative probability of being near the top of the saddle
\cite{HanTalBor90}.  

\begin{figure}
\centering
\includegraphics[width=\medium]{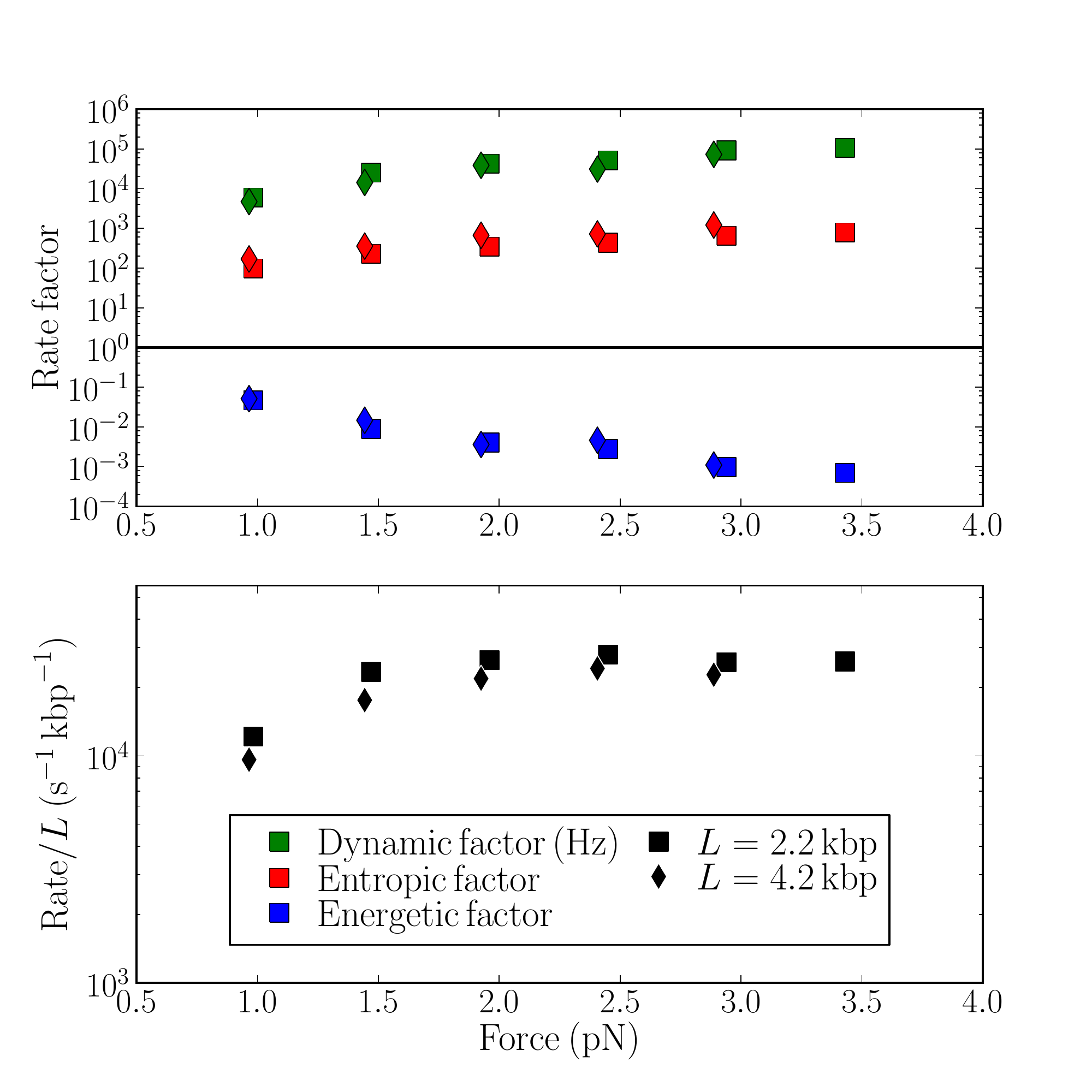}
\caption{\label{RateVsForceFigure}%
(Color online)
Transition state theory plectoneme nucleation  
hopping rate (bottom) and the factors that contribute
to that rate (top) versus external force.  
(Blue; lower) Energetic factor $\exp{(-\Delta E/kT)}$;  
(Red; middle) entropic factor $\exp{(S/k)}$;
(Green; upper) dynamic factor $\lambda_b/2\pi$, in Hz;
(Black) the final hopping rate per unit length from
\eqref{noDisorderRateEqn}.  The calculation is performed
for the experimental conditions in Ref.~\cite{ForDeuShe08},
with $L=2.2$~kbp (squares) and $L=4.2$~kbp (diamonds).
On this log scale, 
adding the three distances from the horizontal line at $10^0$ in 
the top plot produces the final rate.
Note that the entropic factor cancels the slowing from
the energy barrier factor.
(Here the calculation is performed without intrinsic 
bend disorder, producing small or even negative free energy barriers 
--- see section~\ref{hoppingDisorderSection}.)
}
\end{figure}

For the experimental parameters in Table~\ref{parameterTable}, 
using Rotne-Prager dynamics \footnote{
    Using $M^{\mathrm{diagonal}}$ [\eqref{MDiagonal}] produces a 
    comparable unstable mode rate of 
    $\lambda_b/2\pi = 5.6 \times 10^4$~Hz.  We use the more 
    physical $M^{\mathrm{Rotne-Prager}}$ [\eqref{MRotnePrager}] 
    here and in all further calculations.
},
we find 
$\lambda_b/2\pi = 4.0 \times 10^4$~Hz, $\Delta E/kT = 5.5$, and
$S/k = 5.8$, such that $\Delta\F/kT~=~-0.3$.  
There are two 
surprises here: (1) the characteristic rate of motion over the
barrier is very slow compared to typical atomic timescales, and
(2) the entropic factors are so large that they completely
erase the energy barrier.  We consider these issues in more
detail in the next three sections.  We will find that
the slow characteristic
rate comes from the larger length scales involved in the
transition, and that entropy wins over energy due to the length
of the DNA.  (Roughly speaking, since we calculate a 
probability per unit length of a plectoneme critical 
nucleus, for long enough DNA there will always be such a nucleus.)

These rate factors and the corresponding rate per unit length 
of DNA are plotted as a function of
external force for the two experimentally-measured lengths 
in \figref{RateVsForceFigure}  \footnote{
    The fact that the entropy erases the energy barrier means that
    we are not really allowed to calculate a rate in transition 
    state theory.  We will later find, however, that the addition
    of intrinsic bending disorder increases the free energy barrier
    to more reasonable values, and that it does not significantly
    change the rate.  We thus provide the zero-disorder rate as 
    a function of force as an indication of the likely force 
    dependence of the final plectoneme nucleation hopping rate.
}. This rate per unit length itself depends on the length because 
(1) the saddle-state torque $\tau$ changes with $L$ and (2) the 
twisting component of the unstable mode $d\psi$ is length-dependent 
at the lengths of interest \footnote{
    We predict that, in an infinite system, $d\psi$ should 
    die away with a characteristic length of about $3\times 10^3$~nm, 
    but our longest system is shorter than this 
    (about $1.4 \times 10^3$~nm).
}.  

\subsection{Order of magnitude estimates of the dynamical prefactor}
\label{OrderOfMagnitudeSection}


Typically, rates for atomic scale systems have prefactors on the order
of $10^{13}$~Hz.
Indeed, inserting typical atomic length 
scales (\AA) and energy scales (eV) into the simple rate 
equation \eqref{simpleRateEquation}, and using Stokes' law for an
angstrom sized sphere in water, the energy curvature is $1$~eV/\AA$^2$,
and the damping is 
$6\pi \eta r_a = 10^{-13}$~eV~s/\AA$^2$, producing a 
dynamical prefactor of $10^{13}$~Hz.


Why, then, is our dynamical prefactor of order $10^4-10^5$~Hz?
It turns out that the relevant length and energy scales for the 
DNA supercoiling
transition are not atomic.  For the saddle state, we
are dealing with length scales on the order of tens of 
nanometers (much larger than single atoms), and energy scales 
related to the elastic constants 
($B/\ell \approx 10$~pN~nm $< 0.1$~eV).

\begin{figure}
\centering
\includegraphics[width=\medium]{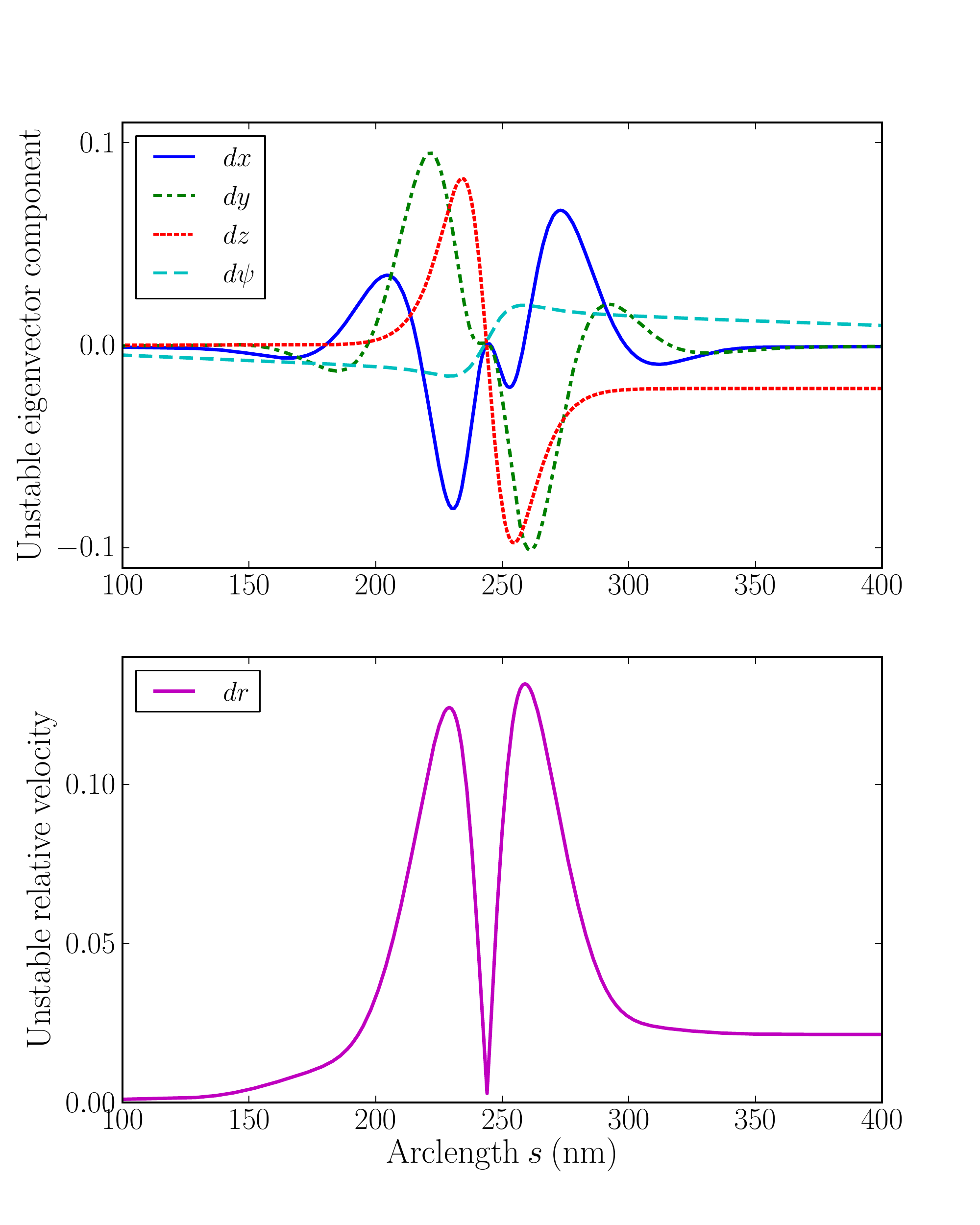}
\caption{(Color online) 
Unstable mode at top of barrier. \label{UnstableModeFigure}%
(top) The four components of the unstable mode eigenvector as a function
of arclength $s$ along the DNA strand.  (bottom) Plot of 
$dr = \sqrt{dx^2+dy^2+dz^2}$; the peaks show the locations where
the (Cartesian) motion of the DNA is greatest when traversing
the barrier.  Note that the width of the peak is about 75 nm ---
inserting this length scale into \eqref{prefactorEstimate} produces 
a prefactor $\sim~10^5$ Hz.
}
\end{figure}

To arrive at a better estimate, we can approximate the saddle
energetics from bending energies only.
Consider approximating the saddle state as a straight
configuration with a single planar sinusoidal bump of 
length $\ell_B$ and amplitude $A$.
Since the elastic bending energy is 
$E_B = \frac{B}{2} \int (\frac{dt}{ds})^2 ds$, where the
relevant component of the
tangent vector is in this case $t = A k_t \cos{( k_t s )}$
for wavenumber $k_t = \pi/\ell_B$, the total bending energy
for the bump is 
$E_B = \frac{B \ell_B}{2} k_t^4 A^2 = \frac{\pi^4 B}{\ell_B^3} \frac{A^2}{2}$;
this leads to an energy curvature with respect to amplitude of
$d^2 E_B / d A^2 = \frac{\pi^4 B}{\ell_B^3}$.  The viscous
damping coefficient corresponding to a rod of radius 
$r_{D}$ and length $\ell_B$ moving sideways through water 
is $\ell_B \zeta$, with $\zeta$ given by \eqref{zetaEqn}.
Putting this together, our back-of-the-envelope estimate
for the prefactor is [see \eqref{simpleRateEquation}]
\begin{equation}
\label{prefactorEstimate}
\frac{\EC}{\Damping} \sim \frac{\pi^4 B}{\ell_B^4 \zeta}.
\end{equation}
We see that the prefactor is strongly dependent on the
length scale $\ell_B$ of the bending of DNA in the unstable 
mode motion.
This length scale should be related to the  
length $\ell \sim 10$~nm, defined in \eqref{ellEqn}, characterizing 
the shape of the saddle configuration.  
As shown in \figref{UnstableModeFigure}, we can check the
amount of DNA involved in the unstable mode motion by looking
at the unstable mode eigenvector.  This reveals that a better
estimate for $\ell_B$ is in fact 75~nm; inserting this into
\eqref{prefactorEstimate} gives a prefactor $\sim 10^5$~Hz,
agreeing with the order of magnitude found in the full 
calculation. This simple calculation shows, then, how 
the 8 orders of magnitude separating the atomic scale rates 
from that of our full DNA calculation arise from the smaller 
energy scales and larger length scales involved.

\subsection{Understanding the entropic factor}

We calculate an entropy $S$ that entirely cancels the energy
barrier $\Delta E$.  What sets the size of $S$?  Comparing 
\eqref{noDisorderRateEqn} and \eqref{effectiveFreeEnergyEqn}, 
we see that the entropic factor
\begin{equation}
e^{(S/k)} = \left(2\pi \frac{L}{\ell_0} J_s J_\rho \right)
    \frac{1}{2\pi kT}\sqrt{ \frac{\det H_{\straight}} {|\det' H_{\sol}|} }.
\end{equation}
This factor comes from comparing the size of fluctuations in 
the normal modes of the straight and saddle states.  We 
expect that most of the modes will be similarly constrained in 
the two states, except for the two zero modes that appear in the
saddle state.  These zero modes create a family of equivalent
saddle points at different locations and rotations 
along the DNA, each of which contributes to the final rate.  
Imagining counting the number of equivalent saddle points 
along $2\pi$ radians in $\rho$ and $L$ nanometers in $s$, 
we can write the entropic factor in the form
\begin{equation}
\label{effectiveEntropyEqn}
e^{(S/k)} = \frac{2\pi}{\rho_0} \frac{L}{s_0},
\end{equation}
where $\rho_0$ and $s_0$ define how far one must move the 
soliton bump along $\rho$ and $s$ to get to an independent saddle 
point \footnote{
    Planck's constant plays an analogous role in
    quantum statistical mechanics.  Also, similar
    quantities $L_0$ and $K_0$ are defined in the
    supplementary material of Ref.~\cite{DanForShe09}.
}.
We expect that $\rho_0$ should be about $\pi$ radians (giving two
independent saddle points at each $s$), and $s_0$ 
should be of the order of the length scale of the soliton,
$\ell \approx 10$~nm, producing $\rho_0 s_0 \sim 30$~nm.
And indeed, using $S/k=5.8$ found in the full calculation,
\eqref{effectiveEntropyEqn} gives $\rho_0 s_0 = 14~$nm \footnote{
    Although this argument explains the order of magnitude 
    of the entropic factor, it does not simply explain 
    the force dependence: we find that $\rho_0 s_0$ has
    a stronger dependence on $F$ than predicted by the 
    dependence of $\ell$ on $F$.
    }.
Thus the size of the entropic factor makes sense: it is large
because there are many equivalent locations along the DNA
where the plectoneme can form.


\subsection{Estimates of the free energy barrier
and bead dynamics}
\label{HistogramSection}

\begin{figure}
\centering
\includegraphics[width=\medium]{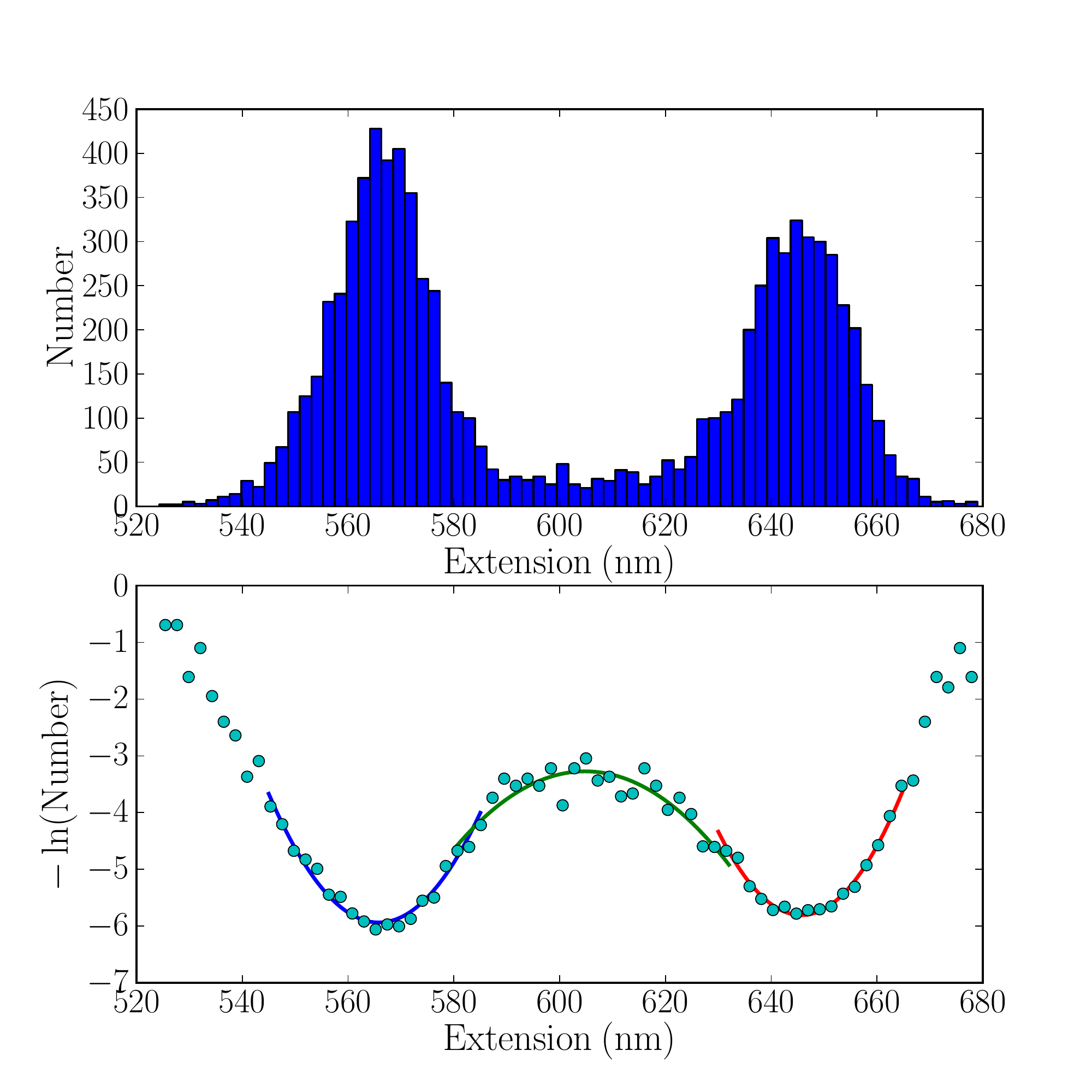}
\caption{ (Color online) 
Bound on free energy barrier from experimental extension 
distribution. \label{HistogramFigure}%
(top) A histogram of measured extensions near the supercoiling
transition, for $F=2$~pN, $L=2.2$~kbp \cite{ForDeuShe08,DanForShe09},
clearly demonstrating bistability.  
(bottom) Fitting the negative natural logarithm of the probability density 
to three quadratic functions
indicates that there is a free energy barrier separating the straight 
state (longer
extensions) from the supercoiled state (shorter extensions) of
at least 2~$kT$.
}
\end{figure}

The effective free energy barrier in our calculation
is reduced to near zero.  This is due to the entropic factor, which 
favors the
saddle state due to the location and rotation zero modes \footnote{
    In the infinite length limit, entropy will always win, smearing
    out the transition.
}.
With a free energy barrier this small, though, plectonemes would form spontaneously
even at zero temperature, and with no barrier to nucleation, no 
bistability would be observed --- indeed, this would violate our 
original assumptions necessary for the use of transition state theory 
itself.
However, the fact that bistability \emph{is} observed in 
experiment \cite{ForDeuShe08,BruLuzKla10}
assures us that the effective free energy barrier is in reality 
nonzero; furthermore, the degree of bistability can give us a reasonable bound
on the size of the barrier.

Two separate experimental groups have directly measured the 
distribution of extensions observed for many seconds near the supercoiling 
transition; one such histogram \cite{ForDeuShe08,DanForShe09} is shown 
in the top of \figref{HistogramFigure}, and the distributions measured 
by the other group \cite{BruLuzKla10} appear 
remarkably similar.  Taking the natural logarithm of this probability density
produces an effective free energy landscape in units of $kT$, shown 
in the bottom of \figref{HistogramFigure}.

To compare the free energy barrier apparent from the extension data to
the one from our calculation [defined in \eqref{effectiveFreeEnergyEqn}],  
there are two subtleties to consider.  First, the measurements with 
extension near the middle, between the two peaks, are not guaranteed to
correspond to configurations that are traversing the saddle between the
two wells (that is to say, extension is not the true reaction coordinate).
Since adding extra probability density unrelated to the transition near 
the saddle point would lower the measured effective free energy barrier, 
we will only be able to put a lower bound on the true 
$\Delta \F$.
Second, looking at the transition state rate equation for one-dimensional
dynamics \cite{HanTalBor90}, 
\begin{equation}
k_{\mathrm{1D}} = \frac{\lambda_b}{2\pi} 
    \sqrt{ \frac{\lambda_{\mathrm{well}}}{\lambda_{\saddle}} } e^{-\Delta E/kT},
\end{equation}
we see the entropic factor coming from the ratio of energy
curvatures in the saddle and well states \footnote{
    If the wells become much narrower than the barrier, they are 
    entropically disfavored, and the barrier-crossing rate increases.
}.
Comparing this to \eqref{effectiveFreeEnergyEqn2}, we see that the
comparable effective free energy barrier should be corrected by this
entropic ratio, such that
\begin{equation}
\Delta \F/kT = \Delta E/kT 
    - \frac{1}{2}\ln{\frac{\lambda_{\mathrm{well}}}{\lambda_{\saddle}}},
\end{equation}
where the $\lambda$s are the curvatures in the well and saddle states.
As shown in the bottom of \figref{HistogramFigure}, we can use fit
parabolas to estimate this entropic correction, finding 0.5~$kT$.
Using the heights of the parabolic fits, we find that $\Delta E = 2.5~kT$,
such that our lower bound on the effective free energy barrier is 
about 2 $kT$.

Finally, we can now check whether the bead dynamics slow the 
hopping measured in experiments.  The curvature of the 
parabola at the top of the barrier in \figref{HistogramFigure}
gives $\lambda_\saddle/2\pi = 3 \times 10^{-3}$~pN/nm;
this matches with the energy curvature $F/L$ that controls
the bead's motion in section~\ref{FullDerivationSection}.
Thus, as in section~\ref{FullDerivationSection}, 
the characteristic rate of bead motion is about 600~Hz ---
using the lower bound of 2 $kT$ for the free energy barrier 
then gives a bead hopping rate of 80 Hz, near the
observed hopping rate.  This provides an explanation for the
discrepancy between the fast rate we calculate and the
slow measured rate: the experimental rate is limited by
the dynamics of the bead.



\section{Including intrinsic bends}
\label{hoppingDisorderSection}

DNA with a random basepair sequence is not perfectly straight, but has intrinsic
bends coming from the slightly different preferred bond angles for each basepair.
This can profoundly affect our calculation by both providing pinning sites for
plectonemes and changing the relevant effective viscosity.

\subsection{History of intrinsic bend measurements}

Since thermal fluctuations also bend DNA, the degree of intrinsic bend disorder is
difficult to measure, but can be estimated using specific DNA sequences that are
intrinsically nearly straight.  The contribution to the bend persistence length
from quenched disorder alone ($P$) can be found using the relation 
\cite{BedFurKat95}
$(B/kT)^{-1} = P^{-1}_{\mathrm{eff}} = P^{-1}_{m} + P^{-1}$:  
$P_{m}$ is found using an intrinsically straight sequence
(such that $P^{-1}=0$) and compared to $P_{\mathrm{eff}}$ from a
random sequence.

Using estimates of wedge angles along with sequence information,
Trifonov et al. estimated $P = 216$ nm \cite{TriTanHar87,Nel98}.
An experiment using cryo-electron microscopy \cite{BedFurKat95}
found $P_m \approx 80$~nm and $P_{\eff} \approx 45$~nm, giving
an intrinsic bend persistence length of $P~\approx~130$~nm.
More recently, a group using  
cyclization efficiency measurements found 
$P_m = 49.5 \pm 1$~nm and $P_{\eff} = 48 \pm 1$~nm, from
which they conclude that
$P > 1000$ nm \cite{VolVol02}, in striking
contrast with the previous estimates.

\begin{figure}
\centering
\includegraphics[width=\medium]{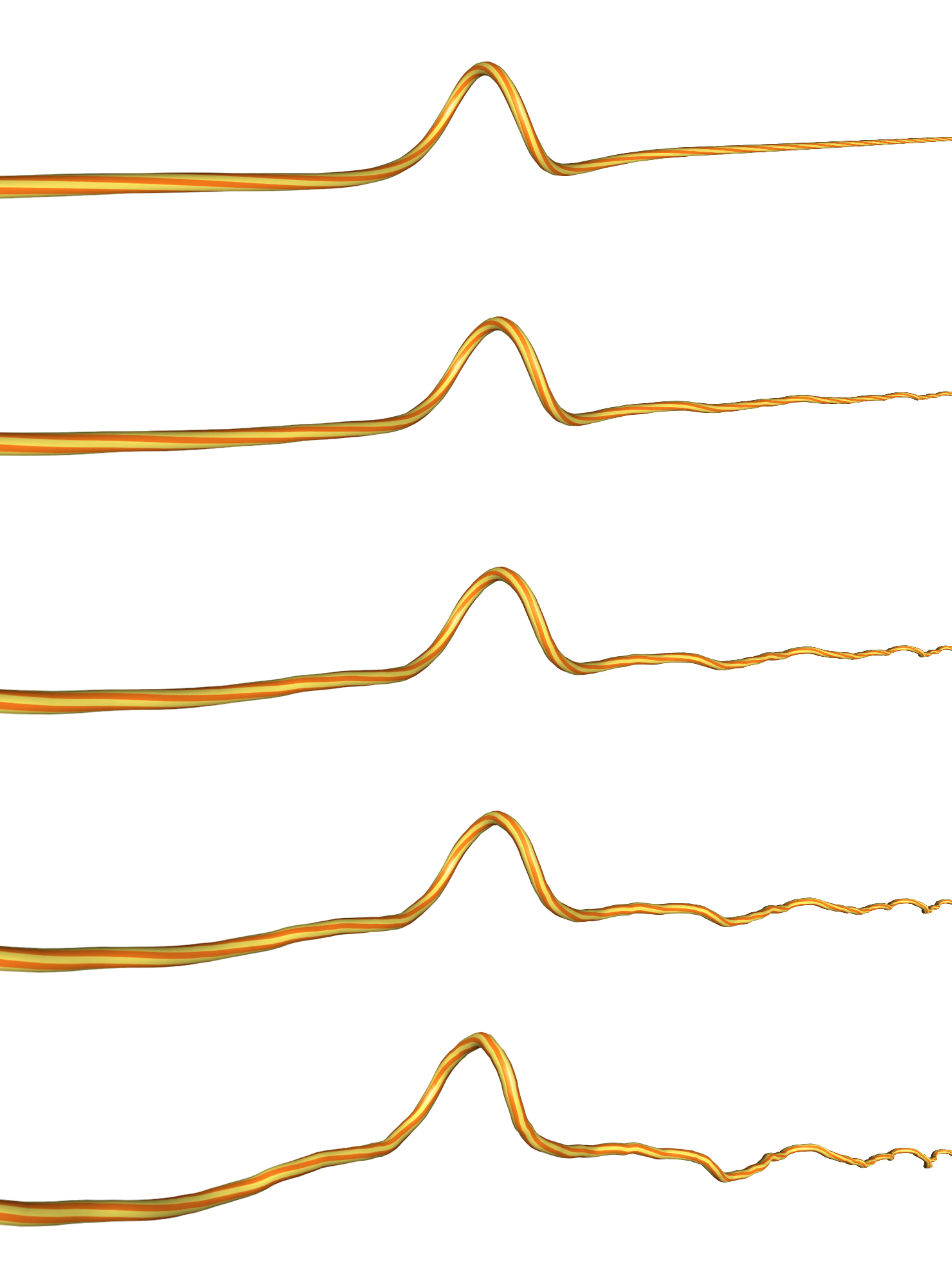}
\caption{ The saddle state with increasing intrinsic bend disorder. 
\label{SaddlePictureTiled}%
(Color online)
From top to bottom, $\D = (0, 0.02, 0.04, 0.06, 0.088)$ nm$^{-1/2}$,
corresponding to persistence lengths $P = (\infty,2500,625,278,130)$ nm,
respectively.  The saddle state is located numerically by 
searching locally for zero force solutions.
}
\end{figure}

We include intrinsic bend disorder in our simulations by shifting
the zero of bending energy for each segment by a random amount,
parameterizing the disorder strength by $D=P^{-2}$ (see 
section~\ref{NumericalDisorderSection} for a detailed description).
We are able to locate the new saddle point including disorder,
as illustrated in \figref{SaddlePictureTiled}, using numerical
methods described in section~\ref{findingSaddlePointsSection}.
Due to the disagreement in the literature about the correct value
of $P$, we treat it as an adjustable parameter and examine
the effects of disorder in a range from
$P = 1000$~nm to $P = 130$~nm ($D = 0.03~\mathrm{nm}^{-1/2}$ to
$D = 0.09~\mathrm{nm}^{-1/2}$).

\subsection{Renormalization of DNA elastic parameters}
\label{renormalizationSection}

It is important to note that the measured elastic constants $B$ and $C$ are
effective parameters that have been renormalized by both thermal
fluctuations and intrinsic bend disorder.  Our simulations do not
explicitly include thermal fluctuations, but incorporate them
by using the measured effective elastic constants.  When we 
explicitly include bend disorder, however, we must use 
microscopic constants $B_m$ and $C_m$ adjusted so they create the 
same large-scale (measured)
effective constants.  Nelson has characterized the first-order
effect of disorder on the elastic constants \cite{Nel98}; 
correspondingly, we use microscopic elastic parameters
\begin{align}
\label{BmEqn}
B_m &= \frac{B}{1-B/(kT P)}, \\
C_m &= C.
\end{align}

%

\subsection{Rate equation with disorder}
\label{rateWithDisorderSection}

If the disorder is large enough, plectoneme formation will be strongly pinned
to one or more locations along the DNA.    
In this case, the zero modes
have vanished, and we find the total rate by adding the contributions from
each saddle point at each location $s_j$. Using \eqref{simpleRateDets},
\begin{equation}
\label{HoppingRateLargeDisorder}
k_\hop = \sum_j \frac{\lambda_{b,j}}{2\pi} 
    \sqrt{ \frac{\det H_{\straight}} {|\det H_{\sol}|} }
    e ^ {-(E_{\sol,j}-E_{\straight})/kT}.
\end{equation}

We can determine when this 
approximation will be valid by checking that fluctuations in the saddle point
position $s_j$ are small compared to the spacing between locations.
The size of fluctuations in $s_j$ can be found in the quadratic approximation as
\begin{align}
\nonumber
\Delta s_j &= \sqrt{ kT \bigg/ \frac{d^2}{d s_j^2}E_{\saddle}(s_j) } \\
    &\approx \sqrt{ kT \bigg/ \frac{d^2}{d s_j^2}\left(\D 
                                        \frac{d E_{\saddle}(s_j)}{d \D} \right) },
    \label{fluctuationsEqn}
\end{align}
where we have replaced $E_{\saddle}$ by its first-order approximation
at low disorder $D$ (see section~\ref{FirstOrderPerturbationSection}).  
Calculating the
second derivative numerically at the pinning sites using \eqref{dEdDEqn}, 
we find
$\frac{d^2}{d s_j^2} \frac{d E_{\saddle}(s_j)}{d \D} \approx 0.5$~pN/nm.
We can thus avoid special treatment of the translation 
modes when the fluctuations in 
$s_j$ are much smaller than the average distance between pinning 
sites, about $75$~nm (see \figref{FirstOrderPerturbationFigureA}).
Setting $\Delta s_j < 75$~nm in \eqref{fluctuationsEqn} then produces
a lower bound on the disorder strength $D$ (or, equivalently, an upper
bound on the intrinsic bend disorder persistence length $P$):
\begin{equation}
\D > 10^{-3}~\mathrm{nm}^{-1/2},~\mathrm{or}~P < 10^6~\mathrm{nm}.
\end{equation}
The experimental estimates for the disorder persistence length $P$
are typically much smaller than this bound (hundreds to thousands
of nanometers), so
the large disorder limit [\eqref{HoppingRateLargeDisorder}] should
be valid for our calculation.

\subsection{Results with disorder}
\label{disorderResultsSection}

\begin{figure}
\centering
\includegraphics[width=\medium]{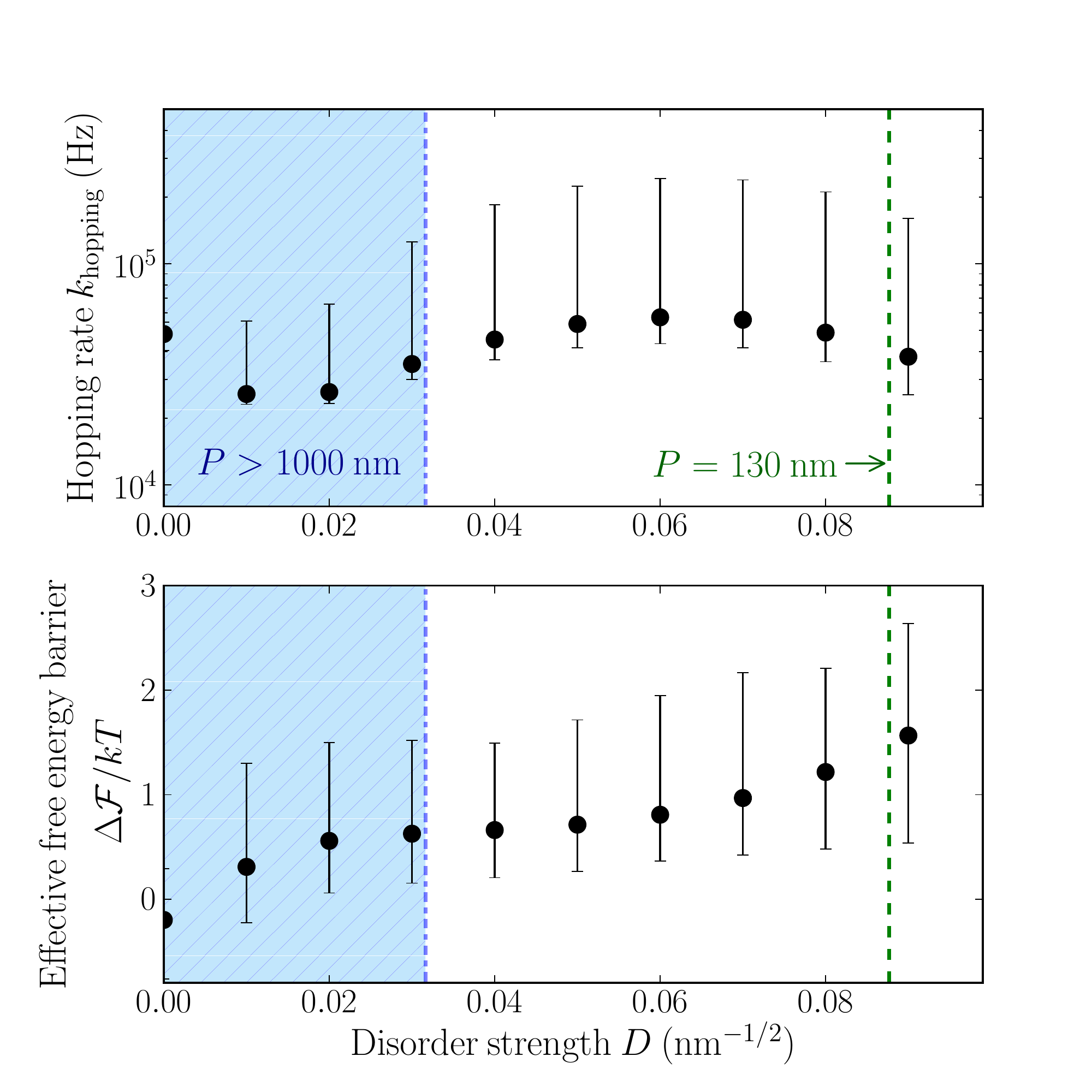}
\caption{Hopping rate and effective free energy barrier 
vs. disorder magnitude, for conditions in Table~\ref{parameterTable}.
\label{DisorderRatePlot}%
(Color online)
The green dashed line corresponds to one experimental 
estimate of the bending order persistence length, $P = 130$~nm 
\cite{BedFurKat95}, and the blue hatched region corresponds to 
another, $P > 1000$~nm \cite{VolVol02}.
The hopping rate $k_{\hop}$ (top) does not change significantly with
the addition of intrinsic bend disorder.   The effective
free energy barrier $\Delta \F$ (bottom) increases with disorder,
such that only the smaller $P$ is consistent with the lower 
bound of 2~$kT$ found using the distribution in \figref{HistogramFigure}.  
The points show results from a single random DNA sequence, and
the error bars are estimates that include uncertainty in DNA sequence
and elastic constants, corresponding to the range of values found 
in \figref{factorsPlot}.
}
\end{figure}

\begin{figure}
\centering
\includegraphics[width=\medium]{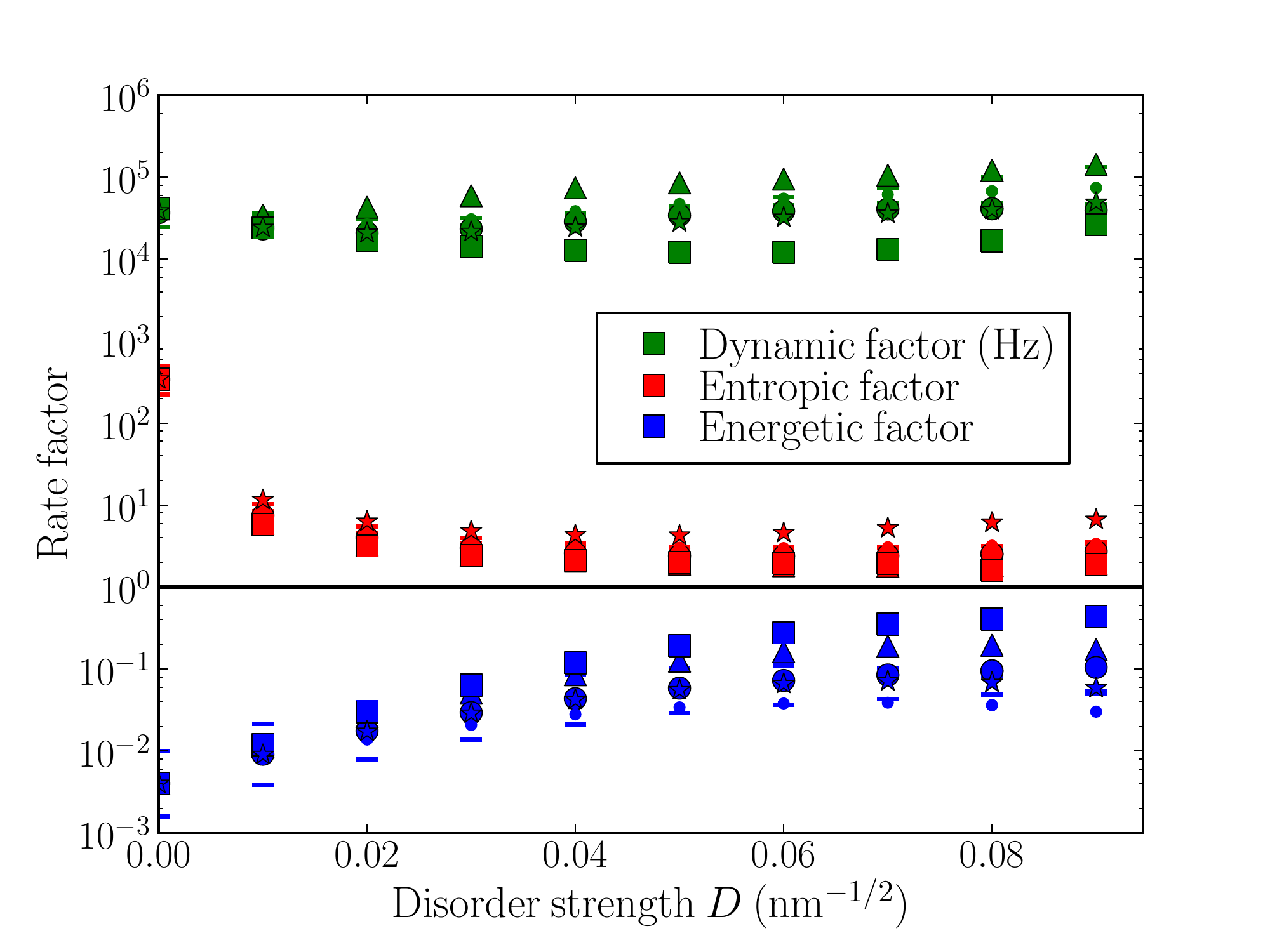}
\caption{
(Color online)
Hopping rate factors versus disorder magnitude for different 
sequences \label{factorsPlot}%
(for the one best plectoneme location $s^*$ when $D>0$, and with
$F=2$~pN, $L=2.2$~kbp).  Colors are
the same as in \figref{RateVsForceFigure}.
Different solid markers correspond to five different
random seeds (five different basepair sequences).  Horizontal bars
correspond to varying the bend elastic constant $B$ for one of the
sequences by adding and subtracting the uncertainty in its measurement, 
(3 nm)$kT$.  On this log scale, 
adding the three distances from the horizontal line at $10^0$ produces 
the final contribution to the rate from location $s^*$. (Including 
multiple plectoneme locations further increases the entropic factor.)  
Both sequence
dependence and parameter uncertainty increase the spread of possible
rates by about an order of magnitude.  
}
\end{figure}

\figref{DisorderRatePlot} displays our results for the hopping 
rate and effective
free energy barrier as a function of the intrinsic bend
disorder strength $D$, for $D$ in the range corresponding to
experimental estimates of the persistence length $P$.
The hopping rate is calculated using \eqref{noDisorderRateEqn}
at zero disorder and \eqref{HoppingRateLargeDisorder} with
disorder (in this case summing over 10 saddle points --- see
section~\ref{findingSaddlePointsSection} for details).  The effective 
free energy barrier is 
calculated using \eqref{effectiveFreeEnergyEqn2} (using the average
$\lambda_b$ over the 10 saddle locations).
We see first that the nucleation rate is not significantly altered by the
intrinsic disorder, remaining between $10^4$ and $10^5$~Hz
(too fast to measure with current experiments).  
The effective free energy barrier, however, rises above zero 
with increasing disorder, making our calculation 
more physically plausible.  Furthermore, note that only the larger 
experimental estimate for intrinsic
disorder ($P=$~130~nm; green dashed line) is consistent with our lower 
bound on $\Delta \F$ of 2~$kT$.

In \figref{factorsPlot}, we plot the components that contribute 
to the rate, defined in \eqref{effectiveFreeEnergyEqn}: the
dynamic factor $\lambda_b$ (green), 
the entropic factor $\exp{(S/k)}$ (red), and the energetic
factor $\exp{(-\Delta E/kT)}$ (blue).  
To explore the variance caused by sequence dependence, we
calculate these factors for five different random intrinsic
bend sequences, for a single plectoneme location $s$ (here, 
the location $s^*$ predicted to have the lowest energy
barrier by first-order perturbation theory; 
see section~\ref{FirstOrderPerturbationSection}).  
Depending on the actual degree of disorder, 
sequence dependence creates a spread in the 
hopping rate of around one order of magnitude.


Since the hopping rate is exponentially sensitive to energy scales at
the transition, it will also be important to carefully consider our knowledge 
of the true elastic constants $B$ and $C$.
Our values [ $B$ = (43 $\pm$ 3 nm)$kT$ and $C$ = (89 $\pm$ 3 nm)$kT$ ] 
were obtained directly from the same experimental setup that produced
the hopping data \cite{ForDeuShe08}, and come from fitting force-extension
data to the worm-like chain model \cite{WanYinLan97}.
The uncertainties in parameters correspond to ranges of rate 
predictions --- 
we numerically check these ranges by performing the rate calculation
using both the upper and lower limits of the ranges for the quoted
value of the two elastic constants.  We find that changes in the elastic
constants mainly affect the rate through the energy barrier $\Delta E$
[\eqref{EnergyBarrierEqn}],
which is much more sensitive to $B$ than to $C$.
The horizontal bars in \figref{factorsPlot} show the results of changing
$B$ from its lower to its upper limit; we see that the uncertainty in 
the bending elastic constant
produces variations on the same scale as the sequence dependence.


\section{Discussion and conclusions}
\label{conclusionSection}

To calculate the rate for plectoneme nucleation at the supercoiling
transition, we first use an elastic rod theory to characterize the saddle
state corresponding to the barrier to hopping.  Using
reaction rate theory, we then calculate the rate prefactor, including
entropic factors and hydrodynamic effects.  We also analyze the
effect of intrinsic bend disorder, which simultaneously lowers 
the energy barrier and increases the entropic barrier.  
We find that the experimental rate 
is in fact set by the slow timescale provided by the bead used to 
manipulate the DNA, with an intrinsic plectoneme nucleation hopping rate 
about 1000 times faster than the measured bead hopping rate.

Further insight  
is gained by studying the factors that contribute to 
the plectoneme nucleation rate.  First, the energy barrier is calculated
analytically using elastic theory [\eqref{EnergyBarrierEqn}].
Second, the rate of motion at the barrier top can be obtained in the
full calculation, and the order of magnitude ($10^5$~Hz) agrees 
with the expected rate of motion of a rod in a viscous fluid when
inserting the appropriate length and energy scales.
Third, the entropic contribution to the prefactor 
significantly lowers the free energy barrier, in
a way directly related to the saddle 
configuration's translational zero mode.  Finally,
from the experimental observations of bistability, we know that
the size of the barrier should be at least $2~kT$ 
(\figref{HistogramFigure}).  

Exploring possible corrections to the calculation, we developed
a method to include disorder due to the randomness in the
basepair sequence.  This disorder introduces a random
intrinsic bend to the DNA, which we are able to incorporate 
by numerically locating the
saddle point configurations.  Intrinsic bends do not significantly
change the hopping rate, though they do increase the
effective free energy barrier (\figref{DisorderRatePlot}).  
Both sequence dependence and uncertainties in the elastic
parameters produce variations in the rate, but they are not
large enough to slow the hopping rate by three orders of 
magnitude to the experimental timescale.   

\begin{figure}
\centering
\includegraphics[width=\medium]{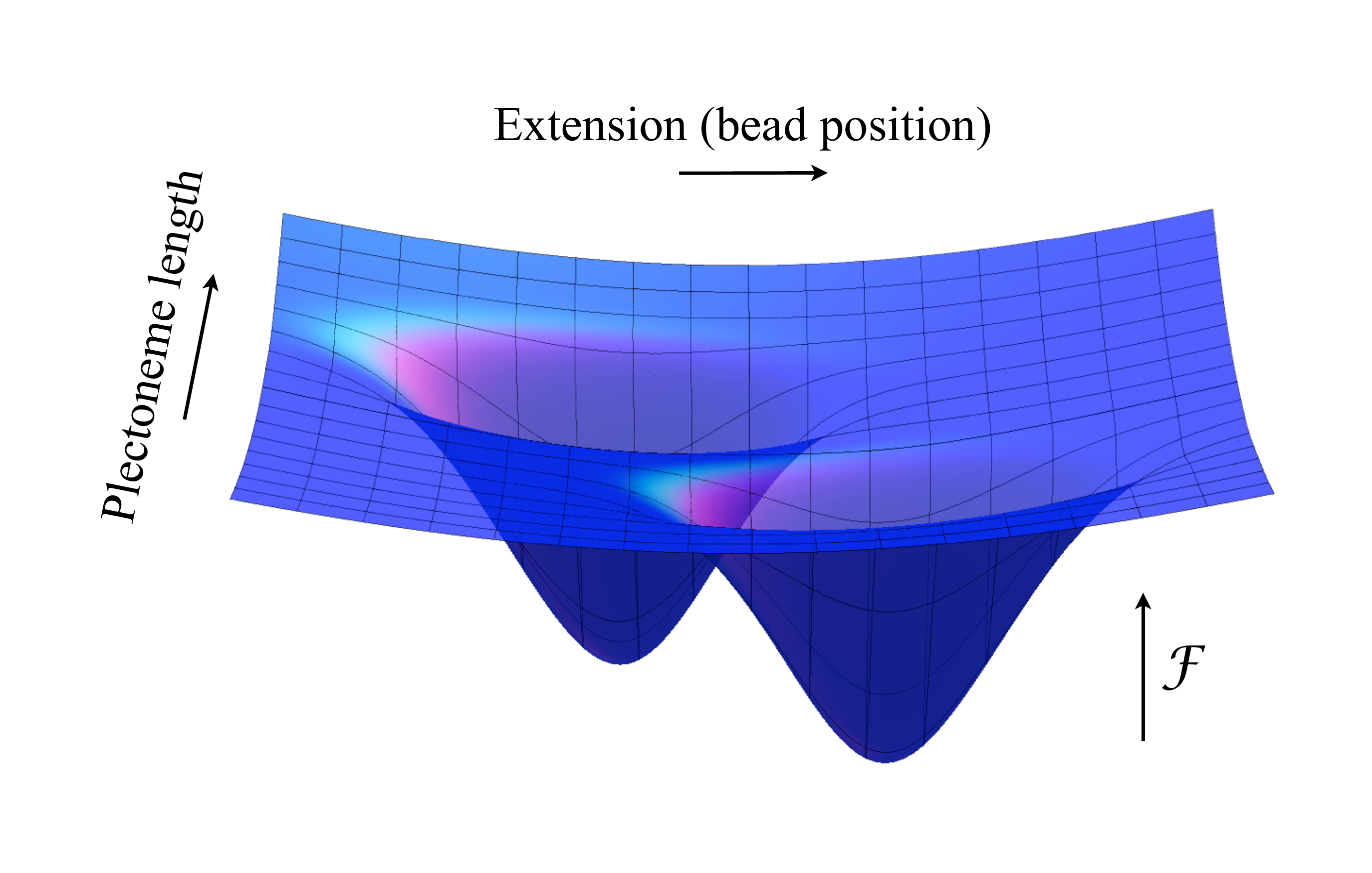}
\caption{\label{FreeEnergySchematic}
(Color online)
Schematic of the double-well in extension and plectoneme length coordinates.
Integrating over the ``plectoneme length'' dimension would produce 
the observed bistable free energy as a function of extension 
shown in \figref{HistogramFigure}.
Note that fixing the
bead position at the saddle extension produces stable states with and
without a plectoneme, with a (smaller) free energy barrier between them, 
producing the fast-timescale hopping behavior illustrated 
in \figref{HoppingSchematic}.}
\end{figure}

\begin{figure}
\centering
\includegraphics[width=\medium]{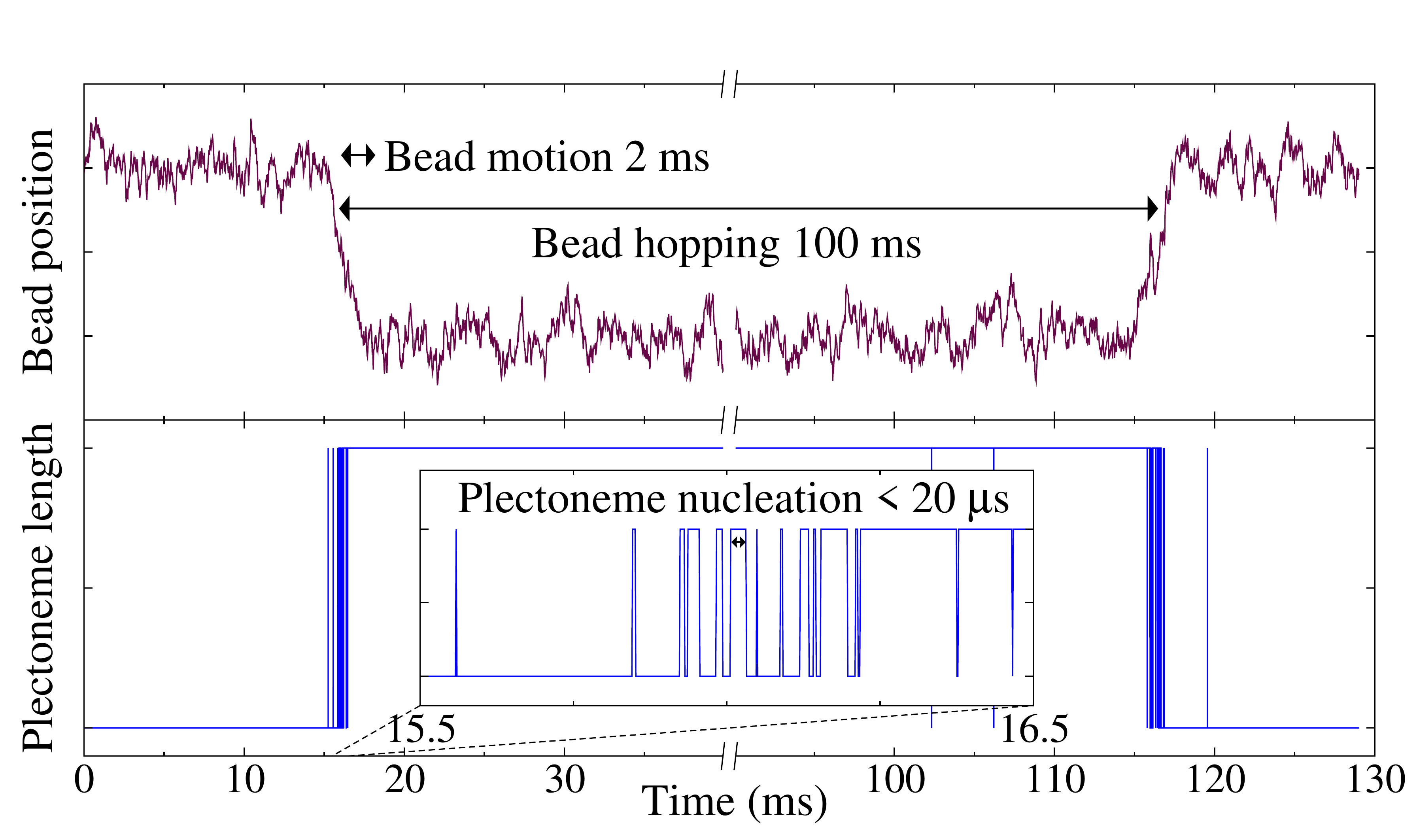}
\caption{
(Color online)
Schematic illustrating relevant timescales.
\label{HoppingSchematic}%
Since the timescale governing the bead's motion through water 
(2~ms) is much slower than the plectoneme nucleation timescale 
(20~$\mu$s), the bead sets the experimentally measured hopping
timescale (100~ms).  
When the bead is located at the saddle point,
the free energy barrier to plectoneme formation is lowered 
(see \figref{FreeEnergySchematic}), such that plectonemes 
nucleate at a rate faster than our calculated
$k_\hop$.  Plectonemes have time to 
form and disappear many times (inset) as the bead 
moves slowly between the two free energy wells.}
\end{figure}

We instead attribute the slowness of the hopping to the
large bead used to manipulate the DNA, since the timescale
controlling the bead's motion ($\omega_b \sim 5 \times 10^2$ Hz) 
is two orders of magnitude slower than the plectoneme nucleation rate 
($k_\hop \sim 5 \times 10^4$ Hz).  
This separation of timescales means that the bead moves through 
an effective free energy potential that is set by all possible 
DNA configurations at a given bead position \footnote{
    One could imagine explicitly calculating this free energy.  Since
    this is both complicated and not biologically relevant, we choose not
    to do so.  Note also that the bead's free energy potential is different
    than the one depicted in \figref{DoubleWellCartoon}.
}.
When the bead is
near the saddle extension, the energy barrier is lower between 
states with and without a plectoneme (see \figref{FreeEnergySchematic}), 
and the microscopic 
configuration of the DNA hops quickly between states with and 
without a plectoneme at a rate faster than $k_\hop$.  But since experiments
measure the bead position, this hopping is invisible, and we see only the slower 
hopping of the bead (of order 10~Hz), which is set by its own viscous drag.
The situation is illustrated in \figref{HoppingSchematic}.

If the characteristic rate controlling the
bead motion were instead made faster than the hopping 
rate, similar experiments could directly measure the plectoneme
nucleation hopping rate.  Some modifications 
would be relatively easy: 
we estimate that 
increasing the external
force $F$ to 5.5~pN (the highest force at which plectonemes are
observed in the current experiments \cite{Maxim}) 
would decrease $k_\hop$ by a factor of 3; 
decreasing the length of the DNA to 1 kbp would
decrease $k_\hop$ by a factor of about 2; reducing the bead's 
size to 100~nm would increase $\omega_b$ by
a factor of 2.  
These modifications would bring the two rates closer by about
one order of magnitude.  We do not see an obvious way to overcome 
the remaining factor of ten, but leave the challenge open
to experimentalists.
If this challenge can be met, future experiments may be 
able to directly measure the plectoneme nucleation
hopping rate, giving useful information about the microscopic dynamics
of DNA in water, and providing a novel testing ground
for transition state theory.


\begin{acknowledgments}
We are grateful for vital insights provided by Michelle Wang, Scott
Forth, and Maxim Sheinin.  Support is acknowledged from NSF grants
DMR-0705167 and DMR-1005479.
\end{acknowledgments}

\appendix




%

\section{The elastic rod model}
\label{elasticRodModel}

The physical properties of long DNA molecules have 
been found to be well-described by linear elastic theory 
(often referred to as the
``worm-like chain'' model, especially in a statistical mechanics context; 
see, e.g., \cite{VolMar97,MarSig95b}).  
In this formulation, the DNA is modeled as a thin elastic rod,
and the energy associated with deforming
it from its natural relaxed state is the sum of local elastic bending,
twisting, and 
stretching energies.   The corresponding elastic constants are sensitive to
experimental conditions such as the ionic concentration of the 
surroundings; in our experimental setup, the bend and stretch
elastic constants $B$ and $S$ can be measured by fitting force-extension curves, 
and
the (renormalized) twist elastic constant $C$ can be measured from the
slope of the torque
as a function of linking number.  These values are listed in
Table~\ref{parameterTable} \cite{ForDeuShe08}.  For the low forces 
in the current experiment
(which are in a biologically-relevant range \cite{ForDeuShe08}), 
the stretch elasticity can be safely ignored \footnote{At the 
    highest force of 3.5 pN and a stretch elastic constant
    of 1200 pN \cite{WanYinLan97}, we expect a strain of 0.3\%, 
    corresponding to an energy
    density of 0.005 pN nm/nm.  This is much smaller than the 
    typical bending and twisting energy densities.
}; we thus treat our DNA as an 
\textit{inextensible} elastic rod, with energy as given in
\eqref{EqElastic}.  Furthermore, since all parts of the DNA stay 
sufficiently far from touching each other in the saddle state,
we also neglect nonlocal repulsive 
interactions (which would be required to stabilize plectonemes), 
allowing an analytical description of the saddle 
state \cite{FaiRudOst96}.

\section{Calculating the energy of a DNA configuration}
\label{energyCalculationSection}

\begin{figure}
\centering
\includegraphics[width=\little]{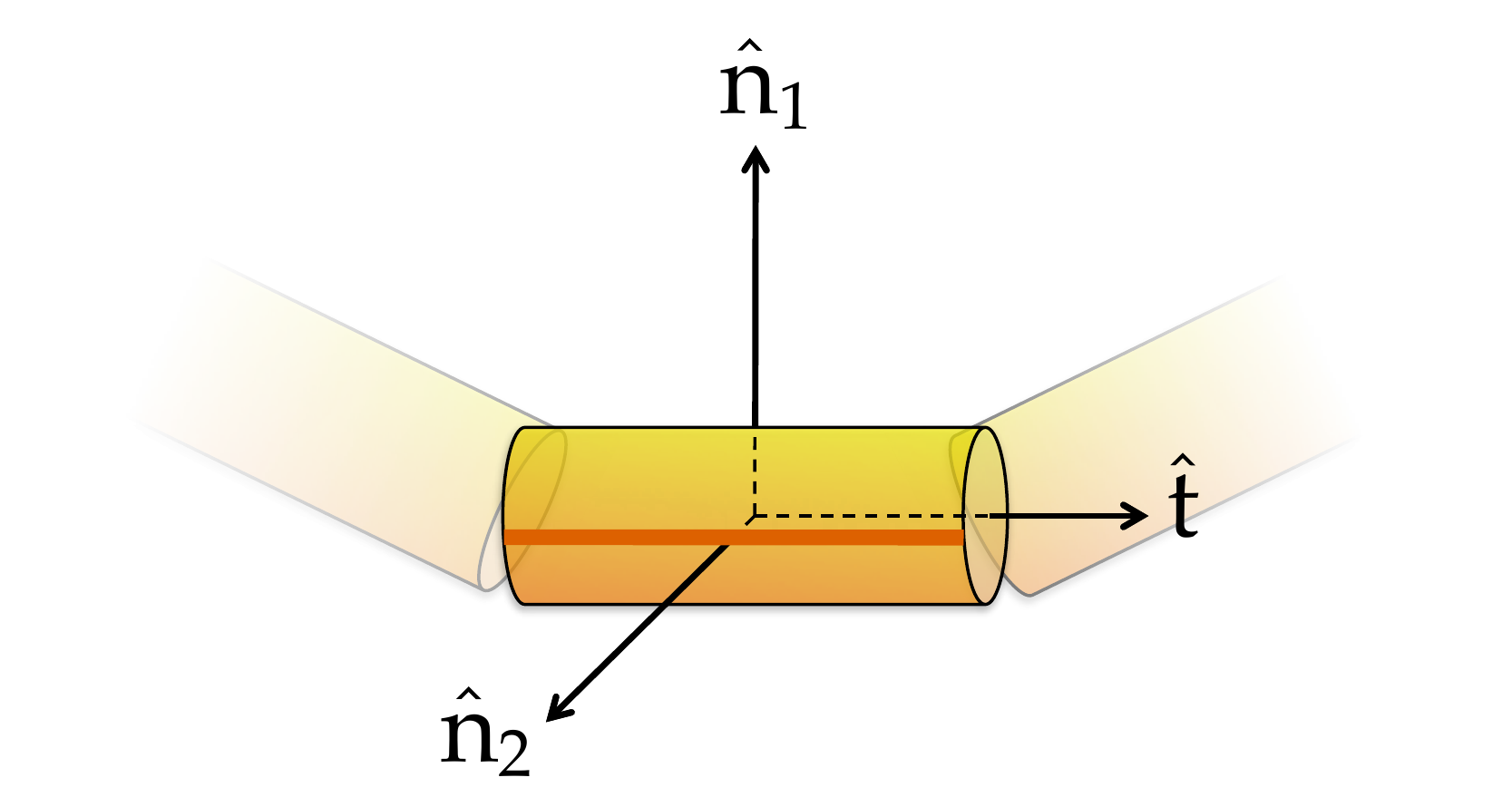}
\caption[]{(Color online) Local basis vectors.\label{LocalBasisFigure}%
}
\end{figure}

In our numerical calculations, we approximate the continuous elastic rod
as a discretized chain of segments, each with fixed length $d$.  
The orientation of 
each segment is described by the rotations necessary to transform the
global Cartesian axes onto the local axes of the segment.  When minimizing
the (free) energy, we find it convenient to use Euler angles \footnote{
We use the `$z$-$x$-$z$' convention: $\psi$ rotates about the original $z$-axis,
$\theta$ rotates about the original $x$-axis, and $\phi$ rotates about the
new $z$-axis.
}
$\phi,\theta,$ and $\psi$, since any set of Euler angles specifies
a valid configuration \footnote{
    Other parameterizations with more degrees of freedom 
    (such as rotation matrices 
    or Cartesian coordinates paired with a local twist) would require 
    constraints to ensure a valid configuration with no stretching.
}.
However, writing
the energy in terms of differences of Euler angles can lead to numerical 
problems near the singularities at the poles.  When calculating energies and
forces, we therefore instead use the full rotation matrix $R$.  $R$ rotates
a segment lying along the $z$-axis to its final orientation; $R$'s
columns are thus the two normal unit vectors followed by the tangent unit 
vector (see \figref{LocalBasisFigure}):
\begin{equation}
R^{(n)} = \left[
\begin{array}{ccc}
\hat n_1^{(n)} & \hat n_2^{(n)} & \hat t^{(n)} 
\end{array} \right]
\end{equation}
\begin{align} \nonumber
\hat n_1 = (& \cos{\phi} \cos{\psi} - \cos{\theta} \sin{\phi} \sin{\psi}, \\ \nonumber
            -&\cos{\psi} \sin{\phi} - \cos{\theta} \cos{\phi} \sin{\psi}, \\ \nonumber
            & \sin{\theta} \sin{\psi} ), \\
\hat n_2 = (& \cos{\theta} \cos{\psi} \sin{\phi} + \cos{\phi} \sin{\psi}, \\ \nonumber
            & \cos{\theta} \cos{\phi} \cos{\psi} - \sin{\phi} \sin{\psi}, \\ \nonumber
            -&\cos{\psi} \sin{\theta} ), \\ \nonumber
\hat t =   (& \sin{\theta} \sin{\phi}, -\cos{\psi} \sin{\theta}, \cos{\theta} ).
\end{align}


There are three degrees of freedom in the ``hinge'' between each segment
that determine the local elastic energy: the two components of bending 
$\beta_1$ and $\beta_2$ (along $\hat n_1$ and $\hat n_2$, respectively), 
and the twist $\Gamma$.   
In terms of the rotation matrix
\begin{equation}
\Delta^{(n)} \equiv (R^{(n)})^T R^{(n+1)},
\end{equation}
which measures the rotation between adjacent segments, 
mapping the $n$th segment's axes 
onto those of the $(n+1)$th,
the bends and twist can be written in 
an explicitly rotation-invariant form: to lowest order in the angles
(see Section~\ref{bendTwistAppendix}),
\begin{align}
\vec \beta \cdot \hat n_1 = \beta_1 &= (\Delta_{23} - \Delta_{32})/2 \\
\vec \beta \cdot \hat n_2 = \beta_2 &= (\Delta_{31} - \Delta_{13})/2 \\
\Gamma &= (\Delta_{12} - \Delta_{21})/2 , \label{GammaEqn}
\end{align}
where the $(n)$ superscripts have been omitted.
The above forms are
useful when the sign of a given component is necessary \footnote{
We will use this,
for example, when we break the symmetry between positive and negative
bends with the introduction of intrinsic bend disorder.
}.
Otherwise, the following squared forms may be used, which have the advantage
of a larger range of validity away from zero 
(see Section~\ref{bendTwistAppendix}):
\begin{align}
\label{betaSqEqn}
\vec \beta^2 &= \beta_1^2 + \beta_2^2 = 2 ( 1 - \Delta_{33} ) \\
\label{xiSqEqn}
\Gamma^2 &= 1 - \Delta_{11} - \Delta_{22} + \Delta_{33}.
\end{align}
Our energy function is then the discretized version of \eqref{EqElastic}:
\begin{equation}
E_{\mathrm{elastic}} = \frac{1}{2d} \sum_m B \vec \beta_m^2 + C\Gamma_m^2, 
\end{equation}
or, separating the bend into its two components,
\begin{equation}
E_{\mathrm{elastic}} = \frac{1}{2d} \sum_m 
    B(\vec \beta_m \cdot \hat n_1^{(m)})^2 
  + B(\vec \beta_m \cdot \hat n_2^{(m)})^2 + C\Gamma_m^2.
\label{bendingEnergyEqn} 
\end{equation}

\section{Transition state calculation details}
\label{transitionStateCalculationSection}

\subsection{Changing to the correct coordinates}
\label{choosingCoordsSection}


So far, we have used Euler angles $(\phi,\theta,\psi)$ to parametrize the 
DNA configuration.  We can easily calculate energy derivatives with respect to
these coordinates. 
However, the hopping rate calculation requires that we do our path integral in 
the same coordinates as the dynamics, given in Section~\ref{DynamicsSection}, 
which are defined in 
Cartesian space with a local twist [$\vec r = (x,y,z,\psi)$].  To efficiently 
calculate the correct Hessian, then, 
we need to convert energy derivatives to $\vec r$ space.

First, we note that there is one less coordinate in Euler angle space: this comes
from the inextensibility constraint, which $\vec r$ space does not have.  Thus
we add a coordinate $\Delta$ to the Euler angles specifying the length of each
segment (which will usually be set to a constant $d$); we will call this
set of coordinates $\vec \alpha = (\phi,\theta,\psi,\Delta)$.
Also, $\vec r$ has $N+1$ elements, each defining the location of one end 
of a segment, while
$\vec \alpha$ has $N$ elements, each defining the Euler angles and length
of each segment.  To match the number of degrees of freedom, we remove
center of mass motion and set constant orientation boundary conditions 
by fixing the location of the first segment's two ends 
and fixing the last segment's orientation along $\hat z$: $\vec r(0) = \vec 0$, 
$\vec r(1) = d \, \hat z$, 
$\vec r(N+1) = \vec r(N) + d \, \hat z$.  This gives a total of $4(N-2)$ 
degrees of freedom.


The Jacobian we would like to calculate is 
\begin{equation}
J_{m i,n j} = \frac{d\vec \alpha_{m i}}{d\vec r_{n j}},
\end{equation}
where $m$ and $n$ label segments and $i$ and $j$ label components.
First writing the Euler angles for a given
segment $n$ in terms of 
$\vec t_n = (x,y,z)_{n+1} - (x,y,z)_{n}$,
\begin{align}
\phi &= \arctan{ \left( -t_y / t_x \right) }                            \\
\theta &= \arccos{\left( t_z / \sqrt{t_x^2 + t_y^2 + t_z^2} \right)}    \\
\Delta &= \sqrt{t_x^2 + t_y^2 + t_z^2}.
\end{align}
We then take derivatives with respect to Cartesian coordinates to produce
the $\phi$, $\theta$, and $\Delta$ rows in the Jacobian.
A subtlety arises in finding expressions for derivatives of the
Euler angle $\psi$ with respect to Cartesian coordinates.
We would like the derivative to correspond to rotating the adjacent
segments to accommodate the change in the location of their connecting
ends.  Due to the way in which Euler angles are defined, this rotation does
not in general leave $\psi$ unchanged, as one might naively expect.
We therefore obtain the derivatives of $\psi$ by first writing the 
rotation matrix corresponding to infinitesimal motion in Cartesian
space and then calculating the corresponding change in $\psi$.  This
produces the nonzero terms in the $\psi$ row of $J_1$ below.

In the end, we have 
\begin{equation}
\label{Jacobian}
J_{mn} = \delta_{m,n} \left[ J_1(m)+J_2 \right] - \delta_{m,n+1} J_1(m),
\end{equation}
where (including the names of the components for 
clarity)
\begin{equation}
\label{J1Eqn}
\begin{array}{ccc}
    ~ & ~ &
    \begin{array}{cccc}
          \;\;\;\;\;\;\;\; x \;\;\;\;\;\;\; & \;\;\;\;\; y \;\; 
        & \;\;\;\;\;\;\;\;\;\; z \; & \;\;\; \psi \;
    \end{array} \\
    J_1(n) \equiv &
    \begin{array}{c}
        \phi    \\
        \theta  \\
        \psi    \\
        \Delta
    \end{array} &
    \left[
    \begin{array}{cccc}
        - t_y / p   & t_y / p     & 0           & 0 \\
        t_x t_z / p \Delta^2      & t_y t_z / p \Delta^2      & -p/\Delta^2 & 0 \\
        t_y t_z / p^2 \Delta      & - t_x t_z / p^2 \Delta    & 0           & 0 \\
        t_x / \Delta              & t_y / \Delta              & t_z/\Delta  & 0
    \end{array}
    \right],
\end{array}
\end{equation}
$p \equiv \sqrt{\Delta^2 - t_z^2}$, $J_2 \equiv \delta_{\psi,\psi}$, and
all the components are evaluated at location $n$.
We use this Jacobian to transform forces with respect to angles $\vec \alpha$
to forces with respect to $\vec r$, which are then used to assemble the
Hessian for use in calculating the unstable mode and the entropic factors
for the transition state calculation in Section~\ref{FullDerivationSection}.


\subsection{Other subtleties}

Since derivatives in $\vec r$ space will in general couple to $\Delta$ (changing
the lengths of segments), we also include an extra stretching energy:
\begin{equation}
E_{\mathrm{stretching}} = \frac{S}{2} \sum_n (\Delta_n - d)^2.
\end{equation}
This avoids problems with extra zero modes corresponding to changing $\Delta$.
We may choose the stretch elastic constant $S$ to match DNA (in which case it
should be about 1000~pN \cite{WanYinLan97}); but since $S$ is so large 
compared to the other elastic constants,
the stretching modes have much higher energy and are the same for the straight
and saddle configurations, canceling in the rate equation [e.g. 
\eqref{simpleRateEigenvalues}].  Thus we find that our results 
are insensitive to the exact value of $S$, as expected.  

As can be seen by inserting $\vec t = (0,0,1)$ into \eqref{J1Eqn}, 
there are singularities in the Jacobian when $\vec t$ points along the 
$z$-axis.  This corresponds to the singularity in the Euler angle representation
at the poles (when $\theta=0$, $\phi$ and $\psi$ are degenerate).  This is
a problem for our formulation because our usual boundary conditions 
hold the ends in the $\hat z$ direction.
As pointed out in Ref.~\cite{KulMohTha07}, a simple way to avoid this
problem is to rotate the system away from the singularity (rotating
the direction of the force as well).  When performing calculations that
require the Jacobian, we therefore rotate the system about the $y$-axis
by an angle $\beta$ and modify the external force term in the energy
from $-F \cos{\theta}$ to 
$-F (\cos{\beta} \cos{\theta} + \sin{\beta} \sin{\theta} \sin{\phi})$.  
This more general formulation also permits an explicit check
that our energies and derivatives are rotation invariant.

The Hessian is constructed by taking numerical derivatives of forces, which
can be calculated analytically.  This gives the Hessian as a 
$4(N-2) \times 4(N-2)$ matrix, which is diagonalized to find eigenvalues
for the entropic calculation [\eqref{simpleRateEigenvalues}].  At zero 
disorder, the zero modes must first be removed; numerically, we find 
that the zero modes show up conveniently as the two modes with smallest
eigenvalues, a few orders of magnitude smaller than any others.


\section{Numerical details}
\label{numericalDetailsSection}

\subsection{Choosing $d$}

\begin{figure}
\centering
\includegraphics[width=\linewidth]{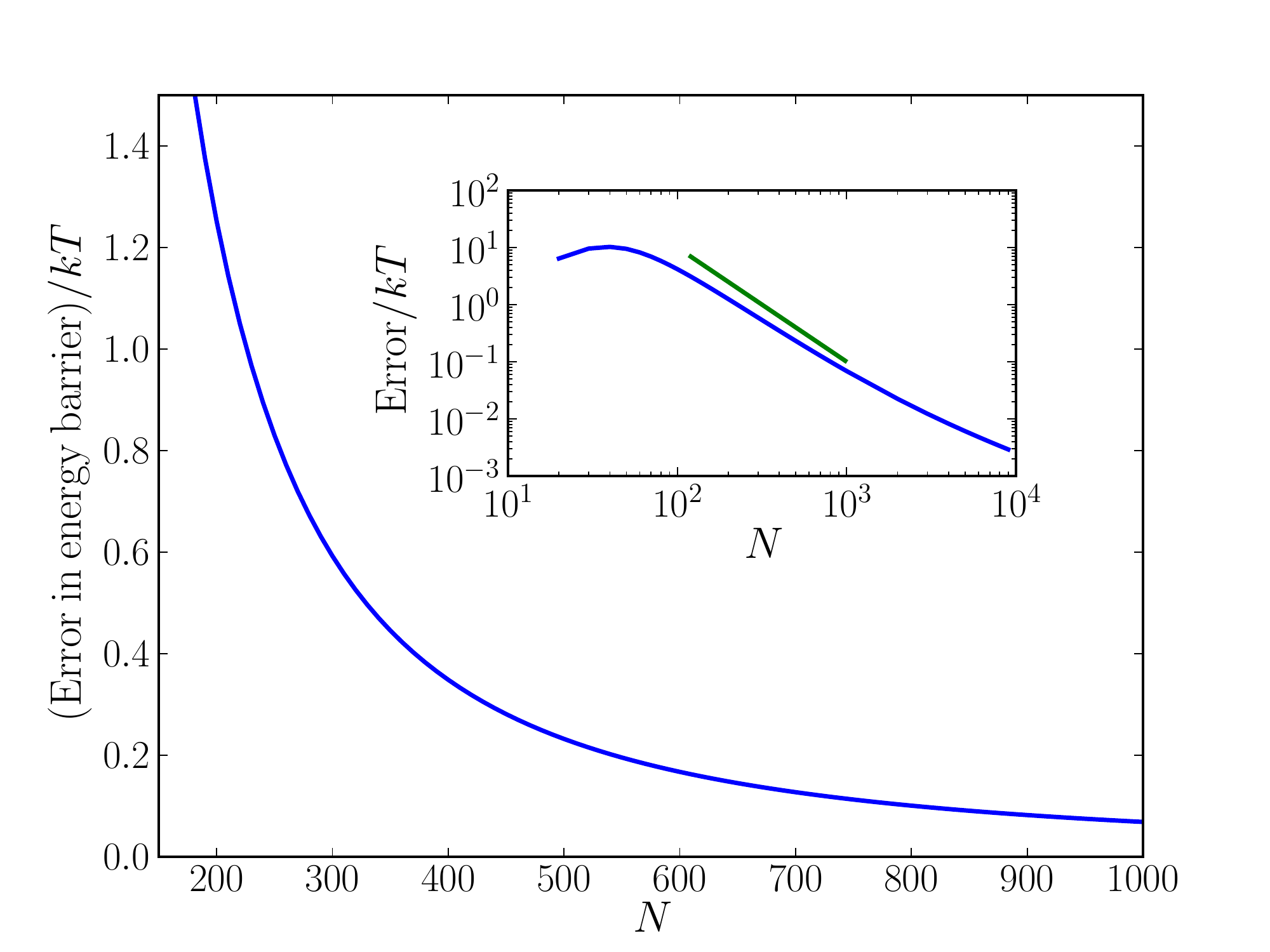}
\caption[]{\label{ChoosingDFigure}%
(Color online)
Choosing $d$. Error in the discretized energy barrier [comparing to the exact result
in \eqref{EnergyBarrierEqn}] as a function of the 
number of discrete segments $N$ (for $L=740$ nm and other parameters
as in Table~\ref{parameterTable}).  Inset: the same data on a log-log plot.  
The green (upper) line 
has a slope of $-2$, showing convergence proportional to $1/N^2$. 
We choose $N=740$ ($d = 1$~nm) as a good trade-off between accuracy
and required computational resources.
}
\end{figure}

As shown in \figref{ChoosingDFigure}, we must be careful to choose
our discretization length (the length $d$ of each segment) such that
our energy calculations are sufficiently accurate.  We can check the
accuracy of the discretized energy calculation by comparing with the 
analytical energy barrier in \eqref{EnergyBarrierEqn}.  
Choosing $d = 1$~nm (about 3 DNA basepairs) produces energy barriers 
within 0.2$kT$ of the 
continuum limit (corresponding to 20\% changes in the hopping rate)
with reasonable memory and time expenditure.

\subsection{Deriving rotation-invariant forms for bend and twist}
\label{bendTwistAppendix}


The amount of local bend and twist can be measured by differences in the rotation
matrices of adjacent segments.  
We would like expressions in terms of the rotation matrices that are correct
to lowest order in the bend or twist angle and that are explicitly rotation invariant.
Our procedure will be to form rotation invariant terms and, writing them
in terms of Euler angles and their derivatives, check what they measure
in terms of bend and twist.

We can first write the bend and twist in terms of derivatives of the local
basis vectors (see \figref{LocalBasisFigure}) or Euler angles 
\cite{FaiRudOst96}:
\begin{align}
\label{BendSquaredEqn}
\beta^2 &= \left[ \dot{\vec t} \, \right]^2 
    = \dot \phi^2 \sin^2 \theta + \dot \theta ^2 \\
\label{TwistSquaredEqn}
\Gamma^2 &= \left[ (\vec n \times \dot{\vec n} ) \cdot \vec t \, \right]^2 
    = \left(\dot \phi \cos \theta + \dot \psi\right) ^2.
\end{align}

We then form rotation-invariant terms from rotation matrices and compare
their Taylor series expansions with respect to bend and twist angles to
\eqref{BendSquaredEqn} and \eqref{TwistSquaredEqn}.
First checking $\mathrm{Tr} [ (R^{(n+1)}-R^{(n)})^T (R^{(n+1)}-R^{(n)}) ]$,
we find that
\begin{align}
\nonumber
\beta^2 + \Gamma^2 &= 
    \frac{1}{2} \left(R^{(n)}_{\alpha \beta}-R^{(n+1)}_{\alpha \beta}\right) 
        \left(R^{(n)}_{\alpha \beta}-R^{(n+1)}_{\alpha \beta}\right) \\
    &= 3 - \Delta_{11} - \Delta_{22} - \Delta_{33}. 
    \label{betaSqPlusXiSqEqn}
\end{align}
Next we check the dot product of the difference in the tangent unit 
vector $\hat t$:
\begin{align}
\nonumber
\beta^2 &= \delta \hat{t} \cdot \delta \hat{t} 
	= \left(R^{(n)}_{\alpha 3}-R^{(n+1)}_{\alpha 3}\right) 
        \left(R^{(n)}_{\alpha 3}-R^{(n+1)}_{\alpha 3}\right) \\
	&= 2 ( 1 - \Delta_{33} ).
    \label{betaSqEqnFull}
\end{align}
\eqref{betaSqPlusXiSqEqn} and \eqref{betaSqEqnFull} produce 
\eqref{betaSqEqn} and \eqref{xiSqEqn}.

To find expressions for signed $\beta$ and $\Gamma$, we notice that the
above use only the diagonal elements of $\Delta$; we can also form rotation 
invariant terms using the off-diagonal elements, producing (with the
Levi-Civita symbol $\varepsilon$)
\begin{align}
\beta_1 &= \varepsilon_{1 \gamma \delta} \Delta_{\gamma \delta} / 2
	= ( \Delta_{23} - \Delta_{32} )/2 \\
\beta_2 &= \varepsilon_{2 \gamma \delta} \Delta_{\gamma \delta} / 2
	= ( \Delta_{31} - \Delta_{13} )/2 \\
\Gamma &= \varepsilon_{3 \gamma \delta} \Delta_{\gamma \delta} / 2
	= ( \Delta_{12} - \Delta_{21} )/2. 
\end{align}

\begin{figure}
\centering
\includegraphics[width=\linewidth]{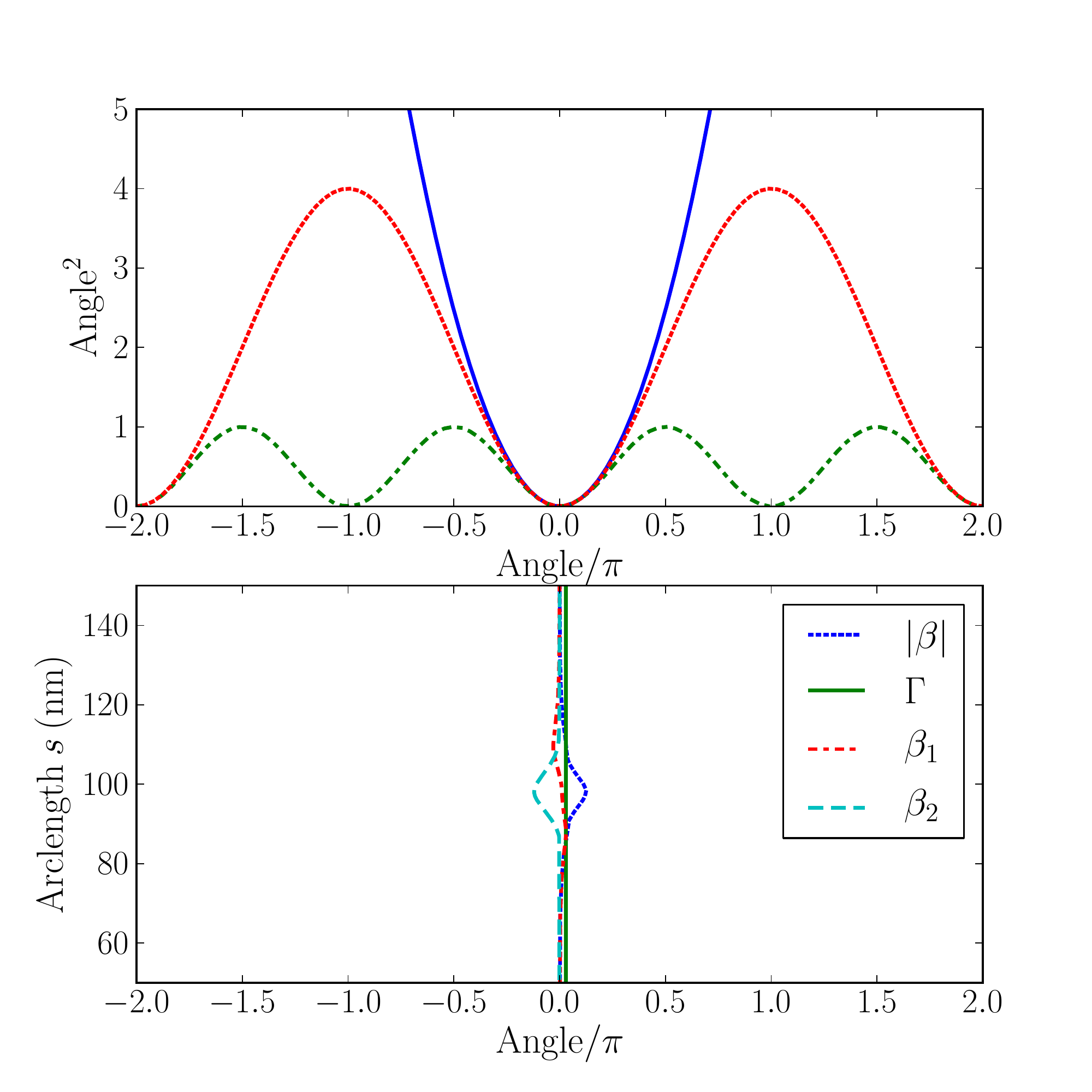}
\caption[]
{
(Color online)
Checking bend and twist expressions. \label{BendTwistMagnitudeFigure}%
(top) Two different rotation-invariant approximations to the bend or twist angle
squared (green, dash-dotted; and red, dotted) compared to the actual 
angle squared (blue, solid).  
Using the non-squared version (green, dash-dotted) leads to a smaller range of
validity.  (bottom) Typical magnitude of bend and twist angles for a plectoneme
(for $F=2$~pN, $d=1$~nm).  Note that the bend and twist angles stay
within the region where either approximation should be valid.
}
\end{figure}

We can see the benefit of using the squared forms 
[\eqref{betaSqEqn} and \eqref{xiSqEqn}] by checking their dependence
on pure bending or twisting rotations.  For example, with 
two segments differing only in twist, 
such that $\psi^{(n+1)} = \psi^{(n)} + \alpha$, we find that
\eqref{GammaEqn} and \eqref{xiSqEqn} produce, respectively \footnote{
    These are straightforwardly evaluated by noting that, e.g., 
    $\Delta_{12} = \hat n_1^{(n)} \cdot \hat n_2^{(n+1)}$.
},
$\Gamma = \sin{\alpha}$ and $\Gamma^2 = 2(1-\cos \alpha)$.  Plotting
$\Gamma^2$ for the two cases (\figref{BendTwistMagnitudeFigure})
demonstrates that they
have the same curvature near $\alpha = 0$ (by design), but 
using the non-squared version in \eqref{GammaEqn} leads to a second minimum  
at $\alpha = \pi$:
we find that, especially when including intrinsic bend disorder, this can
cause the numerical minimizer to  allow $\psi$ to
slip by $\pi$ to the next minimum, unphysically removing linking 
number.  For this reason, we use the squared forms unless
otherwise necessary.

\section{Including intrinsic bend disorder}
\label{disorderAppendix}

\subsection{Setting the disorder size}
\label{NumericalDisorderSection}

As discussed in Section~\ref{hoppingDisorderSection}, we need to include 
DNA's intrinsic bend disorder to understand the energetics of the
saddle point barrier crossing.  To accomplish this, we shift the zero
of the elastic bend energy for each hinge; generalizing \eqref{bendingEnergyEqn},
\begin{equation}
\label{bendingEnergyWithDisorder}
E_{\mathrm{bend}} = \frac{B}{2d} \sum_m 
    ((\vec \beta_m - \sigma_b \vec \xi_m)\cdot \hat n_1^{(m)})^2 
    + ((\vec \beta_m - \sigma_b \vec \xi_m)\cdot \hat n_2^{(m)})^2.
\end{equation}
For each $i$, we choose each of the two components of $\xi$ from a normal 
distribution with unit standard deviation.  We then need to relate 
$\sigma_b$ to the resulting intrinsic bend persistence length $P$.  
The persistence
length is defined by the decay of orientation correlations:
\begin{equation}
\label{persistenceLengthDefinition}
\langle \hat t(0) \cdot \hat t(s) \rangle 
    = \langle \cos \theta(s) \rangle 
    = e^{-s/P},
\end{equation}
where $\theta(s)$ is the angle between segments separated by 
arclength $s$ \cite{BedFurKat95}.  For small $s$ [and thus small
$\theta(s)$], \eqref{persistenceLengthDefinition} becomes
\begin{equation}
\label{PCalculation1Eqn}
1 - \frac{1}{2} \langle \theta(s)^2 \rangle = 1 - \frac{s}{P}.
\end{equation}
At zero temperature and zero force, the size of 
$\langle \theta(s)^2 \rangle$ is set by the intrinsic bends $\xi_m$
only; we are doing a random walk in two dimensions with one step
of root-mean-square size
$\sqrt{2} \sigma_b$ taken for every segment of length $d$.  Thus 
$\langle \theta(s)^2 \rangle = 2 \frac{s}{d} \sigma_b^2$, which
when inserted into \eqref{PCalculation1Eqn} gives the desired relation
between persistence length $P$ and the size of individual random bends
$\sigma_b$:
\begin{equation}
P = \frac{d}{\sigma_b^2}.
\end{equation}
We will also define a convenient parameter $\D$ controlling disorder strength
that is independent of the segment length $d$:
\begin{equation}
\label{Deqn}
\D \equiv \frac{\sigma_b}{\sqrt{d}},
\end{equation}
such that
\begin{equation}
P = \D^{-2}.
\end{equation}


\subsection{First order changes in the energy barrier due to disorder}
\label{FirstOrderPerturbationSection}


\begin{figure}
\centering
\includegraphics[width=\linewidth]{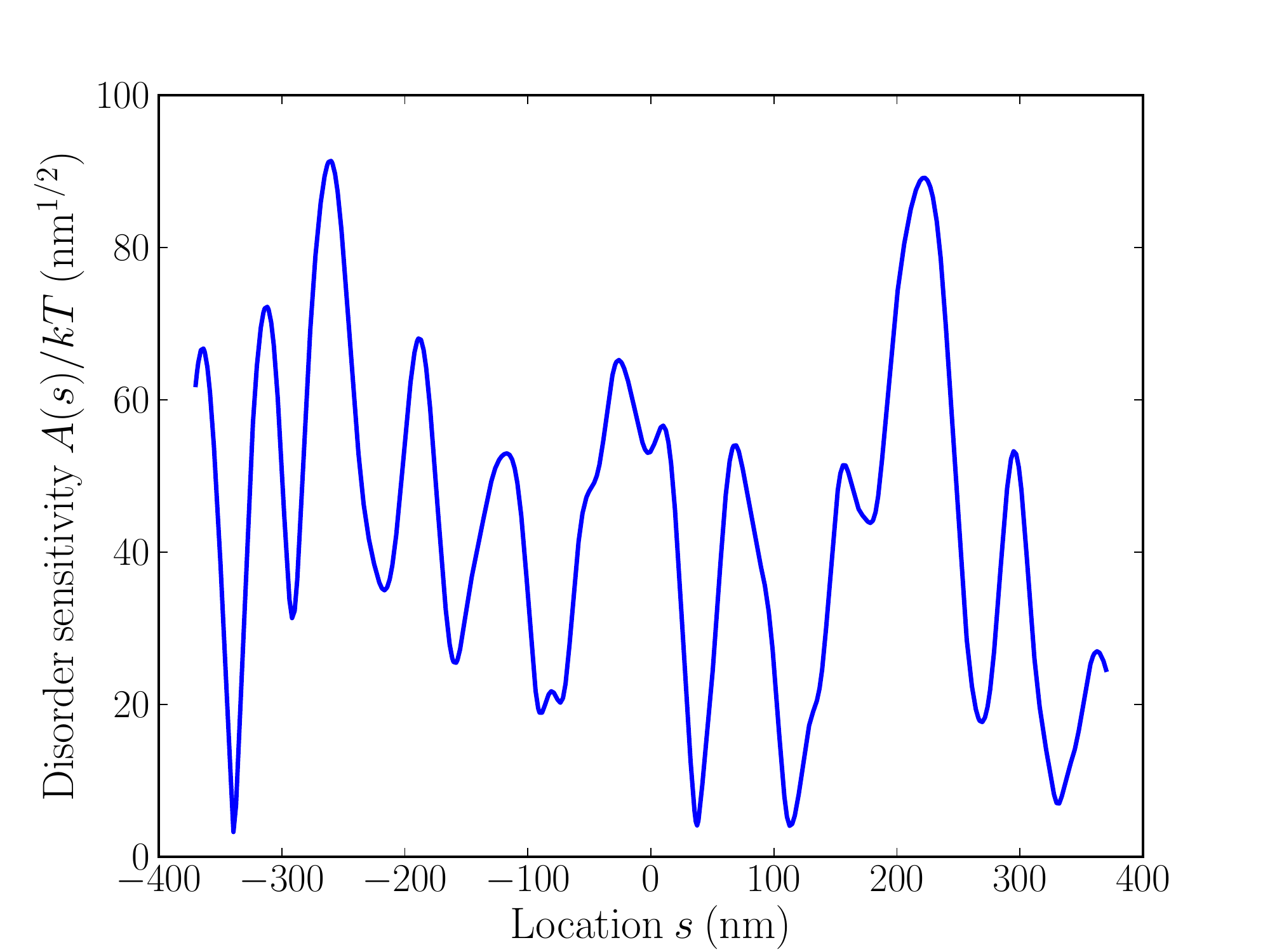}
\caption[]
{
(Color online)
\label{FirstOrderPerturbationFigureA}%
Typical sensitivity of saddle energy to bending disorder strength 
as a function 
of position [$A(s)$, as defined in \eqref{ArhoDefinitionEqn}], 
for $L=740$ nm.  
Peaks indicate positions where plectoneme 
nucleation is energetically favored.  For DNA, the 
bending disorder strength $D$ is estimated to be of order 
0.01 to 0.1~nm$^{-1/2}$; thus we expect pinning to individual sites
(with barriers on the order of $kT$), but also that multiple 
plectoneme locations will contribute to the final hopping
rate in \eqref{HoppingRateLargeDisorder} 
(with multiple locations having energy within about 1 $kT$).
}
\end{figure}

\begin{figure}
\centering
\includegraphics[width=\linewidth]{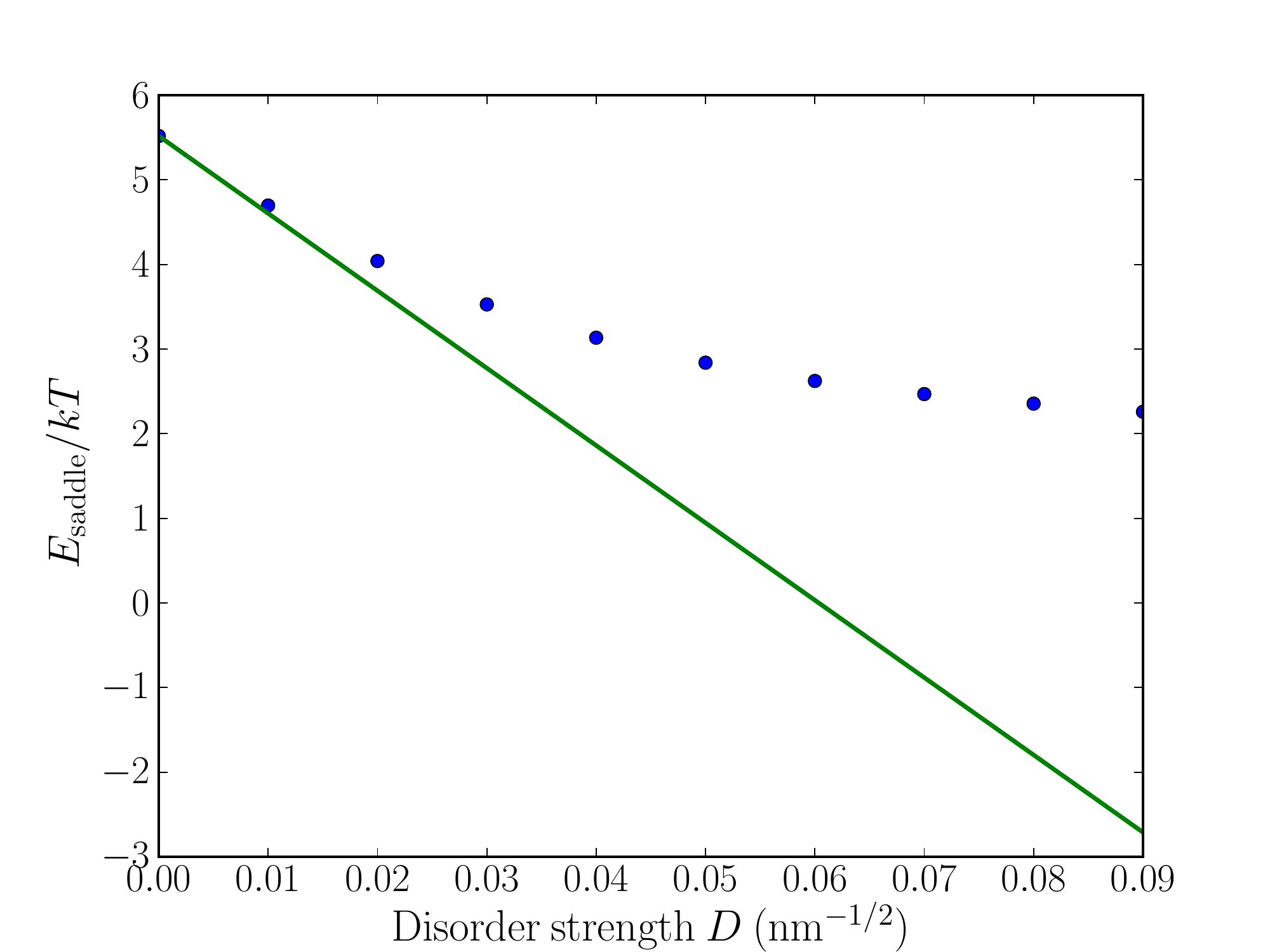}
\caption[]
{
(Color online)
\label{FirstOrderPerturbationFigureESaddle}%
The lowest saddle energy as a function of disorder strength $\D$.  The blue
dots are the true saddle energies, calculated by numerically zeroing forces 
on the saddle configuration.  The green line represents the first-order
approximation given by $D A(s^*)$; we see that the 
approximation correctly predicts the scale of the change, but overestimates 
it by many $kT$ at large disorder.
}
\end{figure}

How does disorder change the energy of the saddle point?  Since the
disorder changes only the bending energy, we can find the lowest
order change from the zero disorder energy by taking the derivative of 
\eqref{bendingEnergyWithDisorder} with respect to
disorder strength $\sigma_b$ at $\sigma_b = 0$:  
\begin{align}
\nonumber
\frac{dE}{d\sigma_b}\bigg|_{\sigma_b=0} 
 &= \frac{dE_{\mathrm{bend}}}{d\sigma_b}\bigg|_{\sigma_b=0} \\
\nonumber
 &= -\frac{B}{d} \sum_m (\vec \beta_m \cdot \hat n_1^{(m)})
                        (\vec \xi_m \cdot \hat n_1^{(m)}) \\
\nonumber
 & \quad \quad \quad \: \: \:  + (\vec \beta_m \cdot \hat n_2^{(m)})
                        (\vec \xi_m \cdot \hat n_2^{(m)}) \\
\label{dEdSigmabEqn}
 &= -\frac{B}{d} \sum_m \vec \beta_m \cdot \vec \xi_m,
\end{align}
or, in terms of disorder strength $\D$ defined in \eqref{Deqn},
\begin{equation}
\label{dEdDEqn}
\frac{dE}{d\D}\bigg|_{\D=0} = 
    -\frac{B}{\sqrt{d}} \sum_m \vec \beta_m \cdot \vec \xi_m.
\end{equation}
We will specifically be interested in the derivative of the saddle
configuration's energy, which will depend on its location $s$ and
rotation $\rho$.  Noting
that $\rho$ will simply rotate the bend 
vector $\vec \beta$, we can write down the form of the 
dependence on $\rho$:
\begin{equation}
\label{ArhoDefinitionEqn}
E_{\saddle}^{'}(s,\rho) \equiv \frac{dE_{\saddle}(s,\rho)}{d\D}\bigg|_{\D=0}
    = -A(s) \cos(\rho - \rho^*(s))
\end{equation}
for some $A(s)$ and $\rho^*(s)$.  
$A(s)$ gives the maximum derivative (sensitivity to disorder) at position $s$, 
and $\rho^*(s)$ is the preferred rotation of the saddle that gives the maximum
(negative) derivative. 
We can find $A(s)$ and $\rho^*(s)$
numerically using the derivative calculated at two values of $\rho$ separated
by $\pi/2$:
\begin{align}
A(s) &= \sqrt{ \left(E_{\saddle}^{'}(s,0)\right)^2 
            +  \left(E_{\saddle}^{'}(s,\pi/2)\right)^2 } \\
\label{phiStarEqn}
\rho^*(s) &= 
    \arctan \left( \frac{E_{\saddle}^{'}(s,\pi/2)}{E_{\saddle}^{'}(s,0)} \right). 
\end{align}
\figref{FirstOrderPerturbationFigureA} shows a typical $A$ as a 
function of $s$.
\figref{FirstOrderPerturbationFigureESaddle} compares the 
first-order approximation to the saddle energy at the location $s^*$, 
given by $D A(s^*)$, to $E_{\saddle}$ calculated numerically by zeroing 
forces (see Section~\ref{findingSaddlePointsSection}).  
We see that the approximation 
correctly predicts the scale of the change for the disorder sizes
in which we are interested, but overestimates the change by many
$kT$ for large disorder.  We therefore use the first-order 
approximation to find the likely locations of the lowest energy
saddle points [at the peaks of $A(s)$], and then zero the forces 
numerically.

To estimate the typical size of this sensitivity 
to disorder, we first note that the bend magnitude for the saddle
configuration is [inserting Eqs.~(\ref{SaddleEulerAngles}) 
into \eqref{BendSquaredEqn}]
\begin{equation}
|\vec \beta_{\saddle}| = \frac{2d}{\ell} \sech{\left( \frac{s}{\ell} \right)}.
\end{equation}
Approximating the function $\sech{(x)}$ as 1 in the range $-2 < x < 2$ and
0 elsewhere, \eqref{dEdDEqn} becomes 
\begin{equation}
\frac{dE_{\saddle}}{d\D}\bigg|_{\D=0} 
    \approx -\frac{B}{\sqrt{d}}\frac{2d}{\ell} \sum_{n=0}^{4\ell/d} \xi_m
\end{equation}
for randomly distributed $\xi$ with unit variance, which produces
\begin{equation}
\sqrt{\big\langle \big(\frac{dE_{\saddle}}{d\D}\big)^2 \big\rangle} 
    \approx \frac{4B}{\sqrt{\ell}}.
\end{equation}
Inserting $B=(43$~nm$)kT$
and $\ell \approx 10$ nm gives a typical sensitivity of about 
$(50$~nm$^{1/2})kT$, agreeing with the scale found in the full 
calculation, as shown in \figref{FirstOrderPerturbationFigureA}.

\subsection{Finding saddle points}
\label{findingSaddlePointsSection}

With large disorder, the saddle points must be found numerically.
We start by estimating the
set of saddle locations $\{s^*\}$ and rotations $\{\rho^*\}$ using first order 
perturbation theory (Section~\ref{FirstOrderPerturbationSection}).  
We find the 
local maxima of $A(s)$ using a one-dimensional local search method starting from
a set of points spaced by the saddle configuration length scale $\ell$.
This gives $\{s^*\}$, from which $\{\rho^*\}$ can be found using \eqref{phiStarEqn}.
For each $s^*$ and corresponding $\rho^*$, we create a zero-disorder saddle 
configuration
[Eqs.~(\ref{SaddleEulerAngles})], and then use this as the starting point 
for a multidimensional 
equation solver (\texttt{scipy.fsolve}) that numerically locates the 
saddle with disorder by finding solutions with zero net force on each 
segment.

\bibliography{thesis}

\begin{thebibliography}{26}
\expandafter\ifx\csname natexlab\endcsname\relax\def\natexlab#1{#1}\fi
\expandafter\ifx\csname bibnamefont\endcsname\relax
  \def\bibnamefont#1{#1}\fi
\expandafter\ifx\csname bibfnamefont\endcsname\relax
  \def\bibfnamefont#1{#1}\fi
\expandafter\ifx\csname citenamefont\endcsname\relax
  \def\citenamefont#1{#1}\fi
\expandafter\ifx\csname url\endcsname\relax
  \def\url#1{\texttt{#1}}\fi
\expandafter\ifx\csname urlprefix\endcsname\relax\def\urlprefix{URL }\fi
\providecommand{\bibinfo}[2]{#2}
\providecommand{\eprint}[2][]{\url{#2}}

\bibitem[{\citenamefont{Forth et~al.}(2008)\citenamefont{Forth, Deufel,
  Sheinin, Daniels, Sethna, and Wang}}]{ForDeuShe08}
\bibinfo{author}{\bibfnamefont{S.}~\bibnamefont{Forth}},
  \bibinfo{author}{\bibfnamefont{C.}~\bibnamefont{Deufel}},
  \bibinfo{author}{\bibfnamefont{M.~Y.} \bibnamefont{Sheinin}},
  \bibinfo{author}{\bibfnamefont{B.}~\bibnamefont{Daniels}},
  \bibinfo{author}{\bibfnamefont{J.~P.} \bibnamefont{Sethna}},
  \bibnamefont{and} \bibinfo{author}{\bibfnamefont{M.~D.} \bibnamefont{Wang}},
  \bibinfo{journal}{Phys. Rev. Lett.} \textbf{\bibinfo{volume}{100}},
  \bibinfo{pages}{148301} (\bibinfo{year}{2008}).

\bibitem[{\citenamefont{Brutzer et~al.}(2010)\citenamefont{Brutzer, Luzzietti,
  Klaue, and Seidel}}]{BruLuzKla10}
\bibinfo{author}{\bibfnamefont{H.}~\bibnamefont{Brutzer}},
  \bibinfo{author}{\bibfnamefont{N.}~\bibnamefont{Luzzietti}},
  \bibinfo{author}{\bibfnamefont{D.}~\bibnamefont{Klaue}}, \bibnamefont{and}
  \bibinfo{author}{\bibfnamefont{R.}~\bibnamefont{Seidel}},
  \bibinfo{journal}{Biophys. J.} \textbf{\bibinfo{volume}{98}},
  \bibinfo{pages}{1267} (\bibinfo{year}{2010}).

\bibitem[{\citenamefont{Bednar et~al.}(1995)\citenamefont{Bednar, Furrer,
  Katritch, Stasiak, Dubochet, and Stasiak}}]{BedFurKat95}
\bibinfo{author}{\bibfnamefont{J.}~\bibnamefont{Bednar}},
  \bibinfo{author}{\bibfnamefont{P.}~\bibnamefont{Furrer}},
  \bibinfo{author}{\bibfnamefont{V.}~\bibnamefont{Katritch}},
  \bibinfo{author}{\bibfnamefont{A.~Z.} \bibnamefont{Stasiak}},
  \bibinfo{author}{\bibfnamefont{J.}~\bibnamefont{Dubochet}}, \bibnamefont{and}
  \bibinfo{author}{\bibfnamefont{A.}~\bibnamefont{Stasiak}},
  \bibinfo{journal}{J. Mol. Biol.} \textbf{\bibinfo{volume}{254}},
  \bibinfo{pages}{579} (\bibinfo{year}{1995}).

\bibitem[{\citenamefont{Trifonov et~al.}(1987)\citenamefont{Trifonov, Tan, and
  Harvey}}]{TriTanHar87}
\bibinfo{author}{\bibfnamefont{E.~N.} \bibnamefont{Trifonov}},
  \bibinfo{author}{\bibfnamefont{R.~K.-Z.} \bibnamefont{Tan}},
  \bibnamefont{and} \bibinfo{author}{\bibfnamefont{S.~C.}
  \bibnamefont{Harvey}}, in \emph{\bibinfo{booktitle}{DNA Bending and
  Curvature}}, edited by \bibinfo{editor}{\bibfnamefont{W.~K.}
  \bibnamefont{Olson}}, \bibinfo{editor}{\bibfnamefont{M.~H.}
  \bibnamefont{Sarma}}, \bibnamefont{and}
  \bibinfo{editor}{\bibfnamefont{M.}~\bibnamefont{Sundaralingam}}
  (\bibinfo{publisher}{Adenine Press}, \bibinfo{year}{1987}), pp.
  \bibinfo{pages}{243--254}.

\bibitem[{\citenamefont{Nelson}(1998)}]{Nel98}
\bibinfo{author}{\bibfnamefont{P.}~\bibnamefont{Nelson}},
  \bibinfo{journal}{Phys. Rev. Lett.} \textbf{\bibinfo{volume}{80}},
  \bibinfo{pages}{5810} (\bibinfo{year}{1998}).

\bibitem[{\citenamefont{Vologodskaia and Vologodskii}(2002)}]{VolVol02}
\bibinfo{author}{\bibfnamefont{M.}~\bibnamefont{Vologodskaia}}
  \bibnamefont{and}
  \bibinfo{author}{\bibfnamefont{A.}~\bibnamefont{Vologodskii}},
  \bibinfo{journal}{J. Mol. Biol.} \textbf{\bibinfo{volume}{317}},
  \bibinfo{pages}{205} (\bibinfo{year}{2002}).

\bibitem[{\citenamefont{Langer}(1968)}]{Lan68}
\bibinfo{author}{\bibfnamefont{J.~S.} \bibnamefont{Langer}},
  \bibinfo{journal}{Phys. Rev. Lett.} \textbf{\bibinfo{volume}{21}},
  \bibinfo{pages}{973} (\bibinfo{year}{1968}).

\bibitem[{\citenamefont{Langer}(1969)}]{Lan69}
\bibinfo{author}{\bibfnamefont{J.~S.} \bibnamefont{Langer}},
  \bibinfo{journal}{Ann. Phys.} \textbf{\bibinfo{volume}{54}},
  \bibinfo{pages}{258} (\bibinfo{year}{1969}).

\bibitem[{\citenamefont{H{\"a}nggi et~al.}(1990)\citenamefont{H{\"a}nggi,
  Talkner, and Borkovec}}]{HanTalBor90}
\bibinfo{author}{\bibfnamefont{P.}~\bibnamefont{H{\"a}nggi}},
  \bibinfo{author}{\bibfnamefont{P.}~\bibnamefont{Talkner}}, \bibnamefont{and}
  \bibinfo{author}{\bibfnamefont{M.}~\bibnamefont{Borkovec}},
  \bibinfo{journal}{Rev. Mod. Phys.} \textbf{\bibinfo{volume}{62}},
  \bibinfo{pages}{251} (\bibinfo{year}{1990}).

\bibitem[{\citenamefont{Coleman}(1977)}]{Col77}
\bibinfo{author}{\bibfnamefont{S.}~\bibnamefont{Coleman}}, in
  \emph{\bibinfo{booktitle}{Proc. Int. School of Subnuclear Physics}}
  (\bibinfo{publisher}{Erice}, \bibinfo{year}{1977}), p. \bibinfo{pages}{265}.

\bibitem[{\citenamefont{Langer and Ambegokar}(1967)}]{LanAmb67}
\bibinfo{author}{\bibfnamefont{J.~S.} \bibnamefont{Langer}} \bibnamefont{and}
  \bibinfo{author}{\bibfnamefont{V.}~\bibnamefont{Ambegokar}},
  \bibinfo{journal}{Phys. Rev.} \textbf{\bibinfo{volume}{164}},
  \bibinfo{pages}{498} (\bibinfo{year}{1967}).

\bibitem[{\citenamefont{McCumber and Halperin}(1970)}]{McCHal70}
\bibinfo{author}{\bibfnamefont{D.~E.} \bibnamefont{McCumber}} \bibnamefont{and}
  \bibinfo{author}{\bibfnamefont{B.~I.} \bibnamefont{Halperin}},
  \bibinfo{journal}{Phys. Rev. B} \textbf{\bibinfo{volume}{1}},
  \bibinfo{pages}{1054} (\bibinfo{year}{1970}).

\bibitem[{\citenamefont{Fain et~al.}(1997)\citenamefont{Fain, Rudnick, and
  {\"O}stlund}}]{FaiRudOst96}
\bibinfo{author}{\bibfnamefont{B.}~\bibnamefont{Fain}},
  \bibinfo{author}{\bibfnamefont{J.}~\bibnamefont{Rudnick}}, \bibnamefont{and}
  \bibinfo{author}{\bibfnamefont{S.}~\bibnamefont{{\"O}stlund}},
  \bibinfo{journal}{Phys. Rev. E} \textbf{\bibinfo{volume}{55}},
  \bibinfo{pages}{7364} (\bibinfo{year}{1997}).

\bibitem[{\citenamefont{Balaeff et~al.}(2004)\citenamefont{Balaeff, Koudella,
  Mahadevan, and Schulten}}]{BalKouMah04}
\bibinfo{author}{\bibfnamefont{A.}~\bibnamefont{Balaeff}},
  \bibinfo{author}{\bibfnamefont{C.~R.} \bibnamefont{Koudella}},
  \bibinfo{author}{\bibfnamefont{L.}~\bibnamefont{Mahadevan}},
  \bibnamefont{and} \bibinfo{author}{\bibfnamefont{K.}~\bibnamefont{Schulten}},
  \bibinfo{journal}{Phil. Trans. R. Soc. Lond. A}
  \textbf{\bibinfo{volume}{362}}, \bibinfo{pages}{1355} (\bibinfo{year}{2004}).

\bibitem[{\citenamefont{Cox}(1970)}]{Cox70}
\bibinfo{author}{\bibfnamefont{R.~G.} \bibnamefont{Cox}}, \bibinfo{journal}{J.
  Fluid Mech.} \textbf{\bibinfo{volume}{44}}, \bibinfo{pages}{791}
  (\bibinfo{year}{1970}).

\bibitem[{\citenamefont{Klenin et~al.}(1998)\citenamefont{Klenin, Merlitz, and
  Langowski}}]{KleMerLan98}
\bibinfo{author}{\bibfnamefont{K.}~\bibnamefont{Klenin}},
  \bibinfo{author}{\bibfnamefont{H.}~\bibnamefont{Merlitz}}, \bibnamefont{and}
  \bibinfo{author}{\bibfnamefont{J.}~\bibnamefont{Langowski}},
  \bibinfo{journal}{Biophys. J.} \textbf{\bibinfo{volume}{74}},
  \bibinfo{pages}{780} (\bibinfo{year}{1998}).

\bibitem[{\citenamefont{M{\"u}lhardt}(2006)}]{Mul06}
\bibinfo{author}{\bibfnamefont{C.}~\bibnamefont{M{\"u}lhardt}},
  \emph{\bibinfo{title}{Molecular Biology and Genomics}}
  (\bibinfo{publisher}{Academic Press}, \bibinfo{year}{2006}).

\bibitem[{\citenamefont{Daniels et~al.}(2009)\citenamefont{Daniels, Forth,
  Sheinin, Wang, and Sethna}}]{DanForShe09}
\bibinfo{author}{\bibfnamefont{B.~C.} \bibnamefont{Daniels}},
  \bibinfo{author}{\bibfnamefont{S.}~\bibnamefont{Forth}},
  \bibinfo{author}{\bibfnamefont{M.~Y.} \bibnamefont{Sheinin}},
  \bibinfo{author}{\bibfnamefont{M.~D.} \bibnamefont{Wang}}, \bibnamefont{and}
  \bibinfo{author}{\bibfnamefont{J.~P.} \bibnamefont{Sethna}},
  \bibinfo{journal}{Phys. Rev. E} \textbf{\bibinfo{volume}{80}},
  \bibinfo{pages}{040901(R)} (\bibinfo{year}{2009}).

\bibitem[{\citenamefont{Wang et~al.}(1997)\citenamefont{Wang, Yin, Landick,
  Gelles, and Block}}]{WanYinLan97}
\bibinfo{author}{\bibfnamefont{M.~D.} \bibnamefont{Wang}},
  \bibinfo{author}{\bibfnamefont{H.}~\bibnamefont{Yin}},
  \bibinfo{author}{\bibfnamefont{R.}~\bibnamefont{Landick}},
  \bibinfo{author}{\bibfnamefont{J.}~\bibnamefont{Gelles}}, \bibnamefont{and}
  \bibinfo{author}{\bibfnamefont{S.~M.} \bibnamefont{Block}},
  \bibinfo{journal}{Biophys. J.} \textbf{\bibinfo{volume}{72}},
  \bibinfo{pages}{1335} (\bibinfo{year}{1997}).

\bibitem[{\citenamefont{Sheinin}()}]{Maxim}
\bibinfo{author}{\bibfnamefont{M.}~\bibnamefont{Sheinin}},
  \bibinfo{note}{personal communication}.

\bibitem[{\citenamefont{Vologodskii and Marko}(1997)}]{VolMar97}
\bibinfo{author}{\bibfnamefont{A.~V.} \bibnamefont{Vologodskii}}
  \bibnamefont{and} \bibinfo{author}{\bibfnamefont{J.~F.} \bibnamefont{Marko}},
  \bibinfo{journal}{Biophys. J.} \textbf{\bibinfo{volume}{73}},
  \bibinfo{pages}{123} (\bibinfo{year}{1997}).

\bibitem[{\citenamefont{Marko and Siggia}(1995)}]{MarSig95b}
\bibinfo{author}{\bibfnamefont{J.~F.} \bibnamefont{Marko}} \bibnamefont{and}
  \bibinfo{author}{\bibfnamefont{E.~D.} \bibnamefont{Siggia}},
  \bibinfo{journal}{Macromolecules} \textbf{\bibinfo{volume}{28}},
  \bibinfo{pages}{8759} (\bibinfo{year}{1995}).

\bibitem[{\citenamefont{Kuli{\'c} et~al.}(2007)\citenamefont{Kuli{\'c},
  Mohrbach, Thaokar, and Schiessel}}]{KulMohTha07}
\bibinfo{author}{\bibfnamefont{I.~M.} \bibnamefont{Kuli{\'c}}},
  \bibinfo{author}{\bibfnamefont{H.}~\bibnamefont{Mohrbach}},
  \bibinfo{author}{\bibfnamefont{R.}~\bibnamefont{Thaokar}}, \bibnamefont{and}
  \bibinfo{author}{\bibfnamefont{H.}~\bibnamefont{Schiessel}},
  \bibinfo{journal}{Phys. Rev. E} \textbf{\bibinfo{volume}{75}},
  \bibinfo{pages}{011913} (\bibinfo{year}{2007}).

\bibitem[{\citenamefont{Crut et~al.}(2007)\citenamefont{Crut, Koster, Seidel,
  Wiggins, and Dekker}}]{CruKosSei07}
\bibinfo{author}{\bibfnamefont{A.}~\bibnamefont{Crut}},
  \bibinfo{author}{\bibfnamefont{D.~A.} \bibnamefont{Koster}},
  \bibinfo{author}{\bibfnamefont{R.}~\bibnamefont{Seidel}},
  \bibinfo{author}{\bibfnamefont{C.~H.} \bibnamefont{Wiggins}},
  \bibnamefont{and} \bibinfo{author}{\bibfnamefont{N.~H.}
  \bibnamefont{Dekker}}, \bibinfo{journal}{Proc. Natl. Acad. Sci. USA}
  \textbf{\bibinfo{volume}{104}}, \bibinfo{pages}{11957}
  (\bibinfo{year}{2007}).

\bibitem[{\citenamefont{Doyle and Underhill}(2005)}]{DoyUnd04}
\bibinfo{author}{\bibfnamefont{P.~S.} \bibnamefont{Doyle}} \bibnamefont{and}
  \bibinfo{author}{\bibfnamefont{P.~T.} \bibnamefont{Underhill}}, in
  \emph{\bibinfo{booktitle}{Handbook of Materials Modeling}}, edited by
  \bibinfo{editor}{\bibfnamefont{S.}~\bibnamefont{Yip}}
  (\bibinfo{publisher}{Springer}, \bibinfo{year}{2005}), p.
  \bibinfo{pages}{2619}.

\bibitem[{\citenamefont{Rotne and Prager}(1969)}]{RotPra69}
\bibinfo{author}{\bibfnamefont{J.}~\bibnamefont{Rotne}} \bibnamefont{and}
  \bibinfo{author}{\bibfnamefont{S.}~\bibnamefont{Prager}},
  \bibinfo{journal}{J. Chem. Phys.} \textbf{\bibinfo{volume}{50}},
  \bibinfo{pages}{4831} (\bibinfo{year}{1969}).

\end{thebibliography}

\end{document}